\documentclass[12pt]{article}
\usepackage{graphicx}
\usepackage{stix}
\usepackage{amsmath,amssymb,amsthm}
\usepackage{bm}
\usepackage{booktabs,multirow}
\usepackage{setspace}
\usepackage{natbib}
\usepackage[dvipsnames]{xcolor}
\usepackage{url}
\usepackage{authblk}
\usepackage{array}
\usepackage{pdfpages}
\usepackage{tabularx}

\definecolor{purple}{rgb}{0.459,0.109,0.538}
\usepackage{geometry}
\def\nat{\mathrm{nat}}
\def\rnt{\mathrm{int}}
\def\sep{\mathrm{sep}}
\def\NIM{\mathrm{NME}}
\def\NDE{\mathrm{NDE}}
\def\NIC{\mathrm{NCE}}
\def\IIM{\mathrm{IME}}
\def\IDE{\mathrm{IDE}}
\def\IIC{\mathrm{ICE}}
\def\SIM{\mathrm{SME}}
\def\SDE{\mathrm{SDE}}
\def\SIC{\mathrm{SCE}}

\newtheorem{assumption}{Assumption}
\newtheorem{theorem}{Theorem}
\newtheorem{lemma}{Lemma}
\newtheorem{remark}{Remark}
\allowdisplaybreaks[4]

\title{\bf Causal mediation analysis for a time-to-event endpoint in the presence of competing events}
\author{Yuhao Deng$^1$, Peter Gilbert$^1$, Youyi Fong$^1$, Bo Zhang$^1$\thanks{Assistant Professor of Biostatistics, Fred Hutchinson Cancer Center. Email: {\tt bzhang3@fredhutch.org}.}}
\affil{\small Vaccine and Infectious Disease Division, Fred Hutchinson Cancer Center \\ 1100 Fairview Ave N, Seattle, WA 98109, USA}
\date{\vspace{-1.4cm}}

\begin{document}

\maketitle
\vspace{-0.3cm}
\begin{abstract}
Causal mediation analysis decomposes a treatment's effect into components operating through distinct causal pathways and characterizes the mediator--outcome relationship. In time-to-event settings, competing events, which preclude the primary endpoint, substantially complicate this decomposition. We provide a unified treatment of causal mediation analysis for a time-to-event endpoint under competing events. We show that the total effect on the cause-specific cumulative incidence decomposes into a direct effect, an indirect effect through the mediator, and an indirect effect through the competing events. We develop this decomposition under the natural, stochastic interventional, and separable effects frameworks, and show that in certain settings all three yield identical statistical functionals despite distinct identification assumptions. We derive efficient influence functions for the counterfactual cumulative incidence curves and the decomposed effects, and propose multiply robust double-machine-learning estimators that attain the semiparametric efficiency bound, together with a sensitivity analysis for unmeasured common causes of the time-to-event endpoints. We further study the mediator--outcome relationship via kernel-smoothed estimands. We apply the methods to immune correlates analyses of COVID-19 vaccine and HIV-1 prevention trials.
\end{abstract}


\section{Introduction}
\label{sec: intro}
\subsection{Causal mediation analysis; applications}
\label{subsec: intro mediation}
Causal mediation analysis elucidates how a treatment's effect operates through post-treatment variables; see, e.g., \citet{imai2010identification}. A prominent application emerged during the COVID-19 pandemic, where multiple mediation frameworks were deployed across a series of phase 3 trials, first for the Ancestral strain \citep{gilbert2022immune, gilbert2022covid} and later for variant strains \citep{zhang2024omicron, zhang2025neutralizing}. The outcome was time to COVID-19, and the mediators comprised post-vaccination immune responses, including antigen-specific neutralizing antibody (nAb) titers and cytokine-releasing T cell proportions.

As in other applications of causal mediation analysis, two questions are of primary interest in these immune correlates studies. The first asks to what extent the vaccine's effect on COVID-19, typically expressed as the causal relative risk or risk difference in the cumulative incidence of COVID-19 up to a prespecified time point (e.g., 6 months post-vaccination), can be attributed to the immune marker of interest. For example, using the method of \citet{benkeser2021inference}, \citet{gilbert2022immune} estimated that neutralizing antibodies mediated approximately two-thirds of the mRNA-1273 vaccine's efficacy in the Coronavirus Efficacy (COVE) phase 3 trial \citep{baden2021efficacy}. Answering this question requires decomposing the total treatment effect into an indirect effect operating through the immune marker of interest and a direct effect operating through other mechanisms. The second question concerns the mediator--outcome relationship: had the vaccine been capable of eliciting immune responses at a given level across the study population, what cumulative COVID-19 incidence would be expected? Using methods developed in \citet{gilbert2023controlled}, \citet[Figure 4B]{gilbert2022immune} depicts the relationship between nAb levels and percentage reduction in COVID-19 risk for five COVID-19 vaccines using phase 3 trial data.

Many conceptual frameworks exist for causal mediation analysis. One strand studies controlled effects \citep{robins1992identifiability, pearl2022direct}, where the quantity of interest is the counterfactual outcome when both treatment and mediator are held fixed at specific levels. Under the natural effects framework, on the contrary, the treatment is set at a fixed level while the mediator takes the value that it would naturally assume under a possibly different treatment level \citep{robins1992identifiability, pearl2014interpretation}. With multiple intermediate variables, natural effects are reformulated as path-specific effects, decomposing the total effect into components operating through distinct causal pathways (see, e.g., \citet{miles2020semiparametric}). The stochastic (or randomized) interventional effects framework instead fixes the treatment while drawing the mediator from a user-specified reference distribution, mimicking an organic data-generating process \citep{vanderweele2014effect, vansteelandt2017interventional, diaz2020causal, Diaz2022Causal}.
Instead of directly manipulating the treatment and mediator, the separable effects framework postulates that treatment comprises two distinct components --- one affecting only the mediator, the other affecting only the outcome --- and that these components could, in principle, be assigned different values in future studies \citep{didelez2019defining, robins2022interventionist, stensrud2022separable}.
Finally, the principal stratification (PS) framework defines subpopulations, termed principal strata, according to the joint potential outcomes of the mediator under treatment and control, and characterizes the treatment effect within these strata \citep{frangakis2002principal, follmann2006augmented, gilbert2008evaluating, vanderweele2008simple, gallop2009mediation}. %Unlike interventionist frameworks, principal stratification remains valid even when the mediator is not manipulable. 

The controlled, natural, stochastic interventional, and separable effects frameworks are all interventionist: each defines effects as contrasts between hypothetical interventions that isolate the pathway through the mediator --- by setting the mediator, letting it take its value under treatment, drawing it from a reference distribution, or assigning a treatment component that acts only through it. This is what allows the total effect to be decomposed into pathway-specific components. The latter three allow us to develop effect decomposition, and the first to study the mediator--outcome dose--response relationship. In contrast, the PS framework characterizes effects within subgroups defined by the mediator’s joint potential values and involves no intervention that isolates the mediator pathway; we do not consider it further in this work.
\subsection{Mediation analysis in the presence of competing events}
\label{subsec: intro lit review and our contribution}

Competing events, defined as events whose occurrence precludes the subsequent occurrence of the primary endpoint in the same individual, can substantially complicate causal mediation analysis. In vaccine studies, there is growing interest in how immune markers mediate a vaccine's effect on time to infection. We focus on the time to first post-vaccination SARS-CoV-2 infection that is asymptomatic, treating symptomatic first infection as a competing event, since its occurrence precludes the first infection from being asymptomatic. An alternative estimand is the time to the asymptomatic first infection, regardless of any prior symptomatic infection. We adopt the former formulation because symptomatic infection is itself a major immunity-conferring event that substantially alters the risk of subsequent asymptomatic infection. %Under our formulation, vaccine efficacy (VE) and its mediation are defined for asymptomatic infection before any such immunity-conferring event has occurred. 

With symptomatic first infection as a competing event, modifying the level of immune response could change the risk of asymptomatic first infection through at least two mechanisms. First, the immune response level may directly affect an at-risk person’s chance of acquiring an asymptomatic infection. Second, the immune response level may affect asymptomatic infection risk indirectly through its influence on symptomatic infection, by (1) converting what would have been symptomatic infections into asymptomatic ones, and (2) preventing symptomatic infections altogether, thereby leaving more people at risk of asymptomatic infection.

Among the most relevant prior work addressing causal mediation analysis with a time-to-event endpoint in the presence of competing events and related problems are \citet{huang2021causal}, \citet{Vo2020Longitudinal}, \citet{domingo2026path}, and \citet{deng2026randomized}. The seminal work of \citet{huang2021causal} studied a setting with an exposure, a primary time-to-event outcome, and an intermediate time-to-event outcome. \citet{huang2021causal} defined potential outcomes for the primary time-to-event endpoint using counting-process notation that depends on treatment assignment and the counting process of the (semi-)competing event, and invoked a pair of sequential ignorability assumptions to identify the indirect effect mediated by the intermediate event and the direct effect. \citet{deng2024direct} discussed sequential ignorability under the natural effects framework, and \citet{martinussen2021estimation} and \citet{breum2024estimation} presented an analog of the dismissible components condition. \citet{Vo2020Longitudinal} considered a setting with a treatment, longitudinal mediators, a primary time-to-event endpoint, and one or more competing events. They studied how to decompose the total effect of treatment on the primary time-to-event outcome into a component operating through longitudinal mediators and another through all survival endpoints within a separable effects framework. They proposed inverse probability weighting-based methods to estimate these two components. \citet{domingo2026path} considered a discrete-time setting with a treatment, a mediator, a primary outcome of interest, and a competing event. They defined three path-specific effects by treating the competing event as an additional mediator, and proposed a regression-based $g$-computation method to estimate them. Finally, \citet{deng2026randomized} estimated stochastic interventional effects using a likelihood-based method in a setting with a time-to-event endpoint, a semi-competing event, and a treatment-induced confounder similar to a mediator.

\subsection{Our contribution}
In this paper, we provide a unified treatment of causal mediation analysis for a time-to-event endpoint in the presence of competing events. We first study effect decomposition under three frameworks: natural effects, stochastic interventional effects, and separable effects frameworks. Unlike prior work, which typically focuses on a single framework, our approach allows us to examine the strengths and limitations of each framework and their connections. We show that the total effect on type-specific risk decomposes into a direct effect on the primary event, an indirect effect through the mediator, and an indirect effect through the competing events. We discuss the identification assumptions underlying each framework and clarify which effects remain identifiable under specific forms of unmeasured confounding. We further show that, in certain settings, the three frameworks yield identical statistical functionals despite relying on distinct identification assumptions. Moreover, unlike prior work that relies on inverse probability weighting alone, outcome regression, or likelihood-based methods, we derive efficient influence functions (EIFs) for the counterfactual cumulative incidence curves and the decomposed effects, and propose EIF-based, double-machine-learning-type estimators. Under mild conditions, the resulting estimators are regular and asymptotically linear, multiply robust, and achieve the semiparametric efficiency bound in a nonparametric model. We then propose a sensitivity analysis to examine the influence of unmeasured common causes across survival endpoints on decomposed effects.

%Second, we study how to estimate the ``dose--response'' relationship between the mediator and the counterfactual risk of the primary event in the presence of competing events. Because the mediator is continuous, the counterfactual risk with the mediator fixed at a prespecified level is not pathwise differentiable in the nonparametric model. We instead target a kernel-smoothed version of this parameter. The smoothed parameter is pathwise differentiable and has a causal interpretation as the counterfactual risk under a stochastic intervention that concentrates the mediator tightly around a prespecified value and generates competing events from a prespecified hazard. We propose an estimator of the kernel-smoothed parameter that is again asymptotically normal and multiply robust. Finally, in Section~\ref{application}, we apply the proposed methods to two immune correlates studies.
Second, we study the dose--response relationship between the mediator and the counterfactual risk of the primary event in the presence of competing events. Because the mediator also alters the hazards of competing events, and hence the population at risk, this risk can be non-monotone in the mediator even when the primary hazard is monotone. We therefore define two controlled-risk estimands, one letting competing events evolve naturally and one fixing their hazards at prespecified levels to isolate the mediator’s effect on the primary event. Because the mediator is continuous, neither estimand is pathwise differentiable \citep{kennedy2017non}. Following \citet{branson2023causal}, we target kernel-smoothed versions, interpretable as counterfactual risks under a stochastic intervention that concentrates the mediator tightly around a prespecified value, and propose asymptotically normal, multiply robust estimators.

%\textcolor{purple}{$\clubsuit$ PG 8-12-26: A good example would have strong surveillance for asymptomatic infection, e.g., many serology tests, or many RNA PCR tests, in all participants (not only symptom triggered diagnostics).$\clubsuit$}

%We also study a useful extension in which the mediator is subject to two-stage sampling and measured only on a subset of study participants — a common design in biomedical studies where measuring the mediator on all participants is too costly. 

\section{Decomposing the effect on the type-specific risk}
\label{sec: decompose total effect}
\vspace{-0.2cm}
\subsection{Notation}
\label{subsec: decompose effect notation}
Let $A$ denote the binary treatment in a two-arm trial, where $A=1$ corresponds to the active treatment and $A=0$ to placebo or an active control. Let $M^a$ denote the potential mediator under treatment $a$. Throughout the development, we let treatment assignment define the time origin. We make the simplifying assumption that the mediator is measured at a fixed time point close to treatment/exposure, and that no failure events (of any competing type) or censoring occur before the mediator is measured. We also assume, for now, that the mediator is measured on all study participants, and we relax this assumption in Section~\ref{subsec: two stage sampling of M}.

%\textcolor{purple}{$\clubsuit$ PG 8-12-26: I think it can't say `at the time origin', because I think mediators have to be measured post treatment/exposure to be eligible to be called a mediator. I think you could just say that the mediator is measured at a fixed time point close to the treatment/exposure, and the simplifying assumptions are made of no failure events (of any of the competing types) and no censoring before the mediator is measured. $\clubsuit$}

For $a \in \{0, 1\}$, let $N^a(t)$ denote the potential counting process under treatment $a$ that jumps at the time of the first event, regardless of type. For $j = 1, \ldots, J$, let $N_j^a(t)$ denote the corresponding potential counting process that jumps at the time of the first event if that event is of type $j$, so that $N^a(t) = \sum_{j=1}^{J} N_j^a(t)$. Throughout the article, we consider only the first event; when it is of type $j$, we refer to it as a type-$j$ first event, or a type-$j$ competing event. We further let $N_0^a(t)$ denote the potential counting process for censoring. Suppose the primary interest is the event type $k\in\{1,\ldots,J\}$. The type-$k$-specific potential cumulative incidence function (CIF) under treatment $a$ is defined as $F_k^a(t) = P\{N_k^a(t)=1\}$. The total effect of $A$ on the type-$k$ first event at time $t$, on the risk difference scale, is defined as the contrast between the potential cumulative incidences under $a = 1$ and $a = 0$:
\begin{equation*}
\tau_k(t) = F_k^{a=1}(t) - F_k^{a=0}(t), \ 0<t<t^*,
\end{equation*}
where $t^\ast$ denotes a prespecified time point that does not exceed the maximum follow-up time.

Our observed data consist of the baseline covariates $X$, the treatment assignment $A$, the mediator $M$, the minimum of the time to the first event or right-censoring $T$, and the event indicator $\Delta$, where $\Delta=j \in \{1,\ldots,J\}$ indicates that the first event was observed and was of type $j$, and $\Delta=0$ indicates right censoring. For $j=1,\ldots,J$, let $N_j(t)=I\{T\le t,\ \Delta=j\}$ denote the observed counting process that jumps at the time of the first event if that event is of type $j$, and let $N_0(t)=I\{T\le t,\ \Delta=0\}$ denote the observed counting process for censoring. The corresponding at-risk process is $Y(t) = I(T \geq t) = I\left\{\sum_{j=0}^{J}N_j(t^-)=0\right\}$, where $t^-$ denotes the instant immediately preceding $t$.

We make the following standard identification assumptions in causal mediation analysis with a time-to-event endpoint \citep{tchetgen2011causal, lange2011direct}.

\begin{assumption}[Standard identification assumptions] \label{ass:std}
(i) Ignorability of treatment assignment: $\{M^{a}, N_j^{a}(\cdot): j=0,\ldots,J\} \perp A \mid X$. (ii) Non-informative censoring: $N_0^{A}(t^-) \perp dN_j^{A}(t) \mid A, X, M^{A}, \sum_{j=1}^{J} N_j^{A}(t^-)=0$. (iii) Positivity: There exists a positive constant $c>0$ such that $P(A \mid X)>c$, $c<f(M^{a'} \mid X)<c^{-1}$ whenever $c<f(M^{a} \mid X)<c^{-1}$, and $P\{\sum_{j=0}^{J} N_j^a(t^*)=0\mid A=a,M^a,X\} > c$ for each $a,a'\in\{0,1\}$. (iv) Consistency: $M^A = M$, $I\left\{\sum_{j=0}^{J}N_j^A(t^-)=0\right\} = Y(t)$, and $dN_j^A(t) \{1-N_0^A(t^-)\} = dN_j(t)$ for each $j\in\{1,\ldots,J\}$ and every $t\in(0,t^*)$.
\end{assumption}

Assumption~\ref{ass:std}(i) states that, conditional on $X$, treatment assignment is independent of all potential outcomes. This assumption holds by design in a randomized trial. Assumption~\ref{ass:std}(ii) states that, among individuals who remain event-free up to time $t$, the occurrence of an event of any type at time $t$ is independent of the censoring process, conditional on $A$, $X$, and $M$. Assumption~\ref{ass:std}(iii) has two implications: (1) the support of $(X, M^a)$ overlaps between the treatment groups, and (2) each treatment group contains individuals who remain at risk throughout the follow-up period. Finally, Assumption~\ref{ass:std}(iv) states that, prior to censoring, the potential mediator and event-counting processes corresponding to the observed treatment assignment coincide with their observed values. Assumption~\ref{ass:std} is sufficient for identifying $\tau_k(t)$, the total effect on the cumulative incidence of each event type. 
%\textcolor{purple}{$\clubsuit$ PG 8-12-26: Assumption 1(i) only holds by the randomization if an intention-to-treat analysis is done, but if the analysis set culls out some subjects (e.g., because they missed vaccinations, or they had an early event, or early dropout), then it won't hold by the randomization. When post treatment mediators are studied, generally the statement is false that it fully holds by the randomization. Yet when combined with your simplifying assumption considered in the previous comment, 1(i) does seem to hold by the randomization. $\clubsuit$} 

The total effect on the type-specific cumulative incidence reflects the combined contributions from (i) the effect mediated through $M$, (ii) the effect operating through competing events, and (iii) the direct effect of treatment not mediated through $M$ or competing events (e.g., via mediators not captured by $M$). To better understand the underlying causal mechanisms, we study how to decompose $\tau_k(t)$ into components corresponding to these three causal pathways.
In the following, we study the decomposition of $\tau_k(t)$ under the natural effects (Section~\ref{sec:nateff}), stochastic interventional effects (Section~\ref{sec:inteff}), and separable effects frameworks (Section~\ref{sec:sepeff}).

\subsection{Natural effects framework} 
\label{sec:nateff}
Decomposing effects within the natural effects framework involves sequential interventions on the treatment, the mediator, and competing events. To formalize these, we begin with potential outcomes under fixed interventions and then extend the notation to random variables and random processes, namely those that would arise naturally under possibly different treatment levels.  Let $dN_j^{a,m}(t)$, $j = 1, \ldots, J$, denote the potential jump of the counting process for the type-$j$ first event in the short time interval $[t,t+dt)$ under treatment $A = a$ and mediator $M = m$. Let $dN_k^{a,m,n_{-k}(t^-)}(t)$ denote the potential jump of the counting process for the type-$k$ first event (the primary endpoint) under treatment $A = a$, mediator $M = m$, and a given status of all competing events immediately before $t$, $n_{-k}(t^-) = \{n_j(t^-): j = 1, \cdots, J,~j\neq k\}$. This notation reflects that whether a type-$k$ first event occurs immediately after time $t$ depends on the assigned treatment, the mediator level, and the realized history of competing events immediately before time $t$. 
For example, in our immune correlates analysis case study, let $j=1$ and $j=2$ denote asymptomatic and symptomatic first infections, respectively. If the endpoint of interest is asymptomatic first infection ($k=1$), then $n_{-1}(t^-) = n_2(t^-)$ is the status of the counting process for symptomatic first infection immediately before time $t$. The potential outcome $dN_1^{a,m,n_{-1}(t^-)}(t)$ then represents whether an asymptomatic first infection occurs in $[t,t+dt)$ under a hypothetical intervention setting vaccination status to $A = a$, the immune marker level $M = m$, and time-$t^-$ symptomatic first infection status to $n_{-1}(t^-)$. If $n_{-1}(t^-)=1$, then we have $dN_1^{a,m,n_{-1}(t^-)}(t)=0$ because only the first event is of interest.

Let $\{n_j(s): 0 \leq s \leq t\}$ denote an intervention that sets the counting process for the type-$j$ first event up to time $t$ to a possible sample path of that process. Let $\widetilde{n}_{-k}(t) = \{\{n_j(s): 0\leq s\leq t\}$, $j = 1, \cdots, J,~j \neq k\}$ collect the interventions of all competing events relative to the type-$k$ first event throughout $[0,t]$. The potential counting process for the type-$k$ first event is then 
\begin{align*}
N_k^{a,m,\widetilde{n}_{-k}(t^-)}(t) &= \int_0^t I\left\{\sum_{j\neq k}n_j(s^-)=0\right\} dN_k^{a,m,n_{-k}(s^-)}(s) = \int_0^t I\left\{\sum_{j\neq k}n_j(s^-)=0\right\} dN_k^{a,m,\overline{0}}(s),
\end{align*}
where $\overline{0}$ is a $(J-1)$-dimensional vector of zeros indicating that no competing event has yet occurred.

Supplementary Figure~S1 illustrates the causal relationships among the treatment, mediator, primary event, and competing events using a directed acyclic graph (DAG) and a single-world intervention graph (SWIG) \citep{richardson2013single}.
%Figure~\ref{fig1}(a) shows the directed acyclic graph (DAG) for a primary event and a competing event, in which the counting process $N_1(t)$ for the primary event depends on both $A$ and $N_2(t^-)$. {\color{purple}Figure~\ref{fig1}(b) shows the corresponding single-world intervention graph (SWIG), obtained by splitting nodes in three steps. First, we split the treatment node into its observed value $A$ and the intervention value $a$, which yields the potential mediator $M^{a}$. Second, we split the mediator node into $M^{a}$ and the intervention value $m$, which yields the potential counting process for the type-2 first event, $N_2^{a,m}(\cdot)$. Finally, we split this node into $N_2^{a,m}(\cdot)$ and the intervention value $n_2(\cdot)$, which yields the potential counting process for the type-1 first event, $N_1^{a,m,n_2(t^-)}(t)$.}
%Figure~\ref{fig1}(b) shows the corresponding single-world intervention graph (SWIG), constructed in three steps. We first set $A=a$ and obtain the potential mediator $M^{a}$, in which we intervene at $m$. We then set $A=a$ and obtain the type-2 event counting process $N_2^{a,m}(\cdot)$, in which we intervene at $n_2(\cdot)$. We finally set $A=a$ and obtain the type-1 event counting process $N_1^{a,m,n_2(t^-)}(t)$.
%\begin{figure}
%    \centering
%    \includegraphics[width=0.9\textwidth]{fig1}
%    \caption{{\small Directed acyclic graph (a) and single-world intervention graph (b) under the natural effects framework (and also the stochastic interventional effects framework).}} \label{fig1}
%\end{figure}

The potential counting processes defined above correspond to fixed interventions on the treatment, the mediator, and the competing-event processes. Under a natural effects framework, the mediator and the competing-event processes are not fixed but take the values they would naturally take under possibly different treatment levels. We now introduce the associated notation. For $a_0\in\{0,1\}$, let $M^{a_0}$ denote the potential mediator under treatment $A=a_0$. For $j=1,\ldots,J$ and $a_j\in\{0,1\}$, let $\{N_j^{a_j,M^{a_0}}(s): 0\le s\le t\}$ denote the potential counting process for the type-$j$ first event under treatment $A=a_j$, with the mediator set to its natural value $M^{a_0}$. We collect the potential counting processes of the competing events relative to the type-$k$ first event as
\[
\widetilde{N}_{-k}^{a_{-k},M^{a_0}}(t)=\Big\{\big\{N_j^{a_j,M^{a_0}}(s): 0\le s\le t\big\}:\ j=1,\ldots,J,\ j\neq k\Big\},
\]
where $a_{-k}=(a_j: j\neq k)$.

Let the $(J+1)$-dimensional vector $\overline{a}=(a_0,a_1,\ldots,a_J)$ denote the treatment levels under which the mediator and each counting process develop naturally; the counterfactual world is indexed by $\overline{a}$. For example, in an immune correlates analysis, the world $\overline{a} = (1,0,0)$ is one in which the immune response $M$ develops naturally under vaccination ($a_0 = 1$), while both asymptomatic and symptomatic first infections develop under withheld vaccination ($a_1 = a_2 = 0$), with the immune response held at its natural level under vaccination.

Finally, equipped with the notation developed above, including $M^{a_0}$, $\widetilde{N}_{-k}^{a_{-k},M^{a_0}}(t)$ and $\overline{a}$, the counterfactual counting process for the type-$k$ first event under $\overline{a} = (a_0, a_1, \cdots, a_J)$ is 
\[
N_k^{\overline{a}}(t) := N_k^{a_k,M^{a_0},\widetilde{N}_{-k}^{a_{-k},M^{a_0}}(t^-)}(t) = \int_0^t I\left\{\int_0^s\sum_{j\neq k}dN_j^{a_j,M^{a_0}}(u^-)=0\right\} dN_k^{a_k,M^{a_0},\overline{0}}(s),
\]
and the corresponding type-$k$-specific counterfactual CIF $F_{k,\nat}^{\overline{a}}(t) = P\{N_k^{\overline{a}}(t)=1\}$, $0<t<t^*$, is of interest under a natural effects framework. 

Identifying $F_{k,\nat}^{\overline{a}}(t)$ requires the following assumption.

%By integrating the counterfactual jumps from $0$ to $t$, we denote the event-$k$ counterfactual counting process under levels $\overline{a}=(a_0,a_1,\ldots,a_J)$ as
%\[
%N_k^{\overline{a}}(t) := N_k^{a_k,M^{a_0},\widetilde{N}_{-k}^{a_{-k}}(t^-)}(t) = \int_0^t I\left\{\int_0^s\sum_{j\neq k}dN_j^{a_j}(u^-)=0\right\} dN_k^{a_k,M^{a_0},\overline{0}}(s),
%\]
%where $\widetilde{N}_{-k}^{a_{-k}}(t) = (N_j^{a_j}(s): j\neq k, s\leq t)$. The primary interest is the type-$k$-specific counterfactual CIF $F_{k,\nat}^{\overline{a}}(t) = P\{N_k^{\overline{a}}(t)=1\}$. The key assumption for identifying natural effects is sequential ignorability.

\begin{assumption} \label{ass:nat}
(i) Ignorability of treatment assignment, $\{M^{a_0},N_k^{a_k,m,\overline{0}}(\cdot),N_j^{a_j,m}(\cdot): j\neq k\}\perp A \mid X$; (ii) no unmeasured confounding between the mediator and survival outcomes, $\{N_k^{a_k,m,\overline{0}}(\cdot), N_j^{a_j,m}(\cdot): j\neq k\} \perp M^{a_0} \mid A=a_0,X$; (iii) no unmeasured confounding between any competing event and other survival outcomes, $\{N_k^{a_k,m,\overline{0}}(t^-), N_j^{a_j,m}(t^-): j\neq k, j\neq l\} \perp dN_l^{a_l,m}(t) \mid N_l^{a_l,m}(t^-)=0,M^{a_l}=m,A=a_l,X$ for every $l\neq k$; (iv) no unmeasured confounding between the primary endpoint and competing events, $\{N_j^{a_j,m}(t^-): j\neq k\} \perp dN_k^{a_k,m,\overline{0}} \mid N_k^{a_k,m,\overline{0}}(t^-)=0,M^{a_k}=m,A=a_k,X$; (v) consistency of potential counting processes, $dN_k^{A,M,N_{-k}(t^-)}(\cdot) = dN_k^{A,M}(\cdot) = dN_k^{A}(\cdot)$, and $dN_j^{A,M}(\cdot) = dN_j^{A}(\cdot)$ for every $j$.
\end{assumption}

Assumption~\ref{ass:nat}(i) strengthens Assumption~\ref{ass:std}(i) by requiring conditional independence between treatment assignment and all versions of potential outcomes. Assumption~\ref{ass:nat}(ii) states that the mediator is independent of potential cross-world event processes within strata defined by covariates and treatment, analogous to the sequential ignorability assumption in classical mediation analysis \citep{joffe2009related, imai2010identification}. Assumptions~\ref{ass:nat}(iii) and (iv) impose a version of sequential ignorability on the jumps of counting processes: conditional on baseline covariates, treatment level, mediator level, and the observed event history, the jumps of counting processes are as-if randomized. Assumptions~\ref{ass:nat}(ii)--(iv) are cross-world assumptions; they are conceptually related to the corresponding assumptions in \citet{deng2024direct}, although that work instead formulated their assumptions in terms of cross-world conditional probabilities. Our setup excludes treatment-induced common causes of mediator and subsequent event processes that arise before mediator measurement. Apart from the mediator and the observed event history, we do not allow additional post-treatment variables to confound relations among the event processes. Finally, Assumption~\ref{ass:nat}(v) is a standard consistency assumption. 

Let $\lambda_j(t;a,m,x) := \lim_{dt\to0} P(t\leq T<t+dt, \Delta=j \mid T\geq t, A=a, M=m, X=x)/dt$ denote the identifiable type-$j$-specific hazard for $j=1,\ldots,J$, and  $\Lambda_j(t;a,m,x) = \int_0^t \lambda_j(s;a,m,x)ds$ denote the cumulative type-$j$-specific hazard. Theorem \ref{thm:nat} provides the key identification result for $F_{j,\nat}^{\overline{a}}(t)$ under Assumptions \ref{ass:std} and \ref{ass:nat}.

\begin{theorem} \label{thm:nat}
Under Assumptions \ref{ass:std} and \ref{ass:nat}, the counterfactual type-$k$-specific CIF $F_{k,\nat}^{\overline{a}}(t) = P\{N_k^{\overline{a}}(t)=1\}$ is identifiable for $0<t<t^*$ and $\overline{a} = (a_0, a_1, \ldots, a_J) \in \{0,1\}^{J+1}$ as follows: 
\begin{equation}
\label{eq: survival g-formula}
    F_{k,\nat}^{\overline{a}}(t) = \int_{\mathcal{X}} \int_{\mathcal{M}} \int_0^t \exp\left\{-\sum_{j=1}^{J}\Lambda_j(s;a_j,m,x)\right\} d\Lambda_k(s;a_k,m,x) dP(m;a_0,x) dP(x),
\end{equation}
where $\mathcal{X}$ is the support of $X$, $P(x)$ is the probability measure of $X$, $\mathcal{M}$ is the support of $M$, and $P(m;a,x)$ is the probability measure of $M$ conditional on $A=a$ and $X=x$.
\end{theorem}

%We will refer to Equation \eqref{eq: survival g-formula} as the survival $g$-formula. 

We observe that $F_k^a(t) = F_{k,\nat}^{(a,a,\ldots,a)}(t)$ for $a\in\{0,1\}$, so the total effect $\tau_k(t)$ admits the following decomposition into a natural effect through the mediator (NME), a natural direct effect (NDE) on the type-$k$ first event, and a natural effect through competing events (NCE):
\begin{align}
\label{eq: decomp nat effect}
&\tau_k(t) := F_k^{a=1}(t) - F_k^{a=0}(t) 
= \NIM_k(t) + \NDE_k(t) + \NIC_k(t),
\end{align}
where $\NIM_k(t) = F_{k,\nat}^{\bm{e}_0}(t) - F_{k,\nat}^{\overline{0}}(t)$, $\NDE_k(t) = F_{k,\nat}^{\bm{e}_k}(t) - F_{k,\nat}^{\bm{e}_0}(t)$, $\NIC_k(t)= F_{k,\nat}^{\overline{1}}(t) - F_{k,\nat}^{\bm{e}_k}(t)$,
with $\overline{1}=(1,\ldots,1)$, $\overline{0}=(0,\ldots,0)$, $\bm{e}_0=(1,0,\ldots,0)$, and $\bm{e}_k$ a length-$(J+1)$ vector whose first and $(k+1)$-th elements equal 1 and all other elements equal 0. This decomposition follows chronological order, reflecting that treatment first affects the mediator and then affects the survival outcomes over time. In our running example of immune correlates analysis, the vaccine's total effect on the cumulative incidence of asymptomatic first infection can be decomposed into (i) $\NIM_1$ ($\overline{a} = (1,0,0)$ versus $(0,0,0)$), the portion attributable to vaccine-induced changes in the immune marker; (ii) $\NDE_1$ ($\overline{a} = (1,1,0)$ versus $(1,0,0)$), the portion attributable to the direct influence of the vaccine on asymptomatic first infection risk; and (iii) $\NIC_1$ ($\overline{a} = (1,1,1)$ versus $(1,1,0)$), the portion attributable to the direct influence of the vaccine on symptomatic first infection. This decomposition requires the comparator-arm hazard to be evaluated at marker values attained under the investigational vaccine, so the positivity condition in Assumption~\ref{ass:std}(iii) is essential. The total effect can alternatively be decomposed as $\text{NME} + \text{NCE}' + \text{NDE}'$, with $\text{NCE}'$ comparing $\overline a=(1,0,1)$ with $(1,0,0)$ and $\text{NDE}'$ comparing $(1,1,1)$ with $(1,0,1)$. The mediation effect is the same in both; the two decompositions differ only by the interaction $I = F^{(1,1,1)} - F^{(1,1,0)} - F^{(1,0,1)} + F^{(1,0,0)}$, with $\text{NDE}' = \text{NDE} + I$ and $\text{NCE}' = \text{NCE} - I$, which vanishes when the primary- and competing-event components act additively on the CIF. Reporting both decompositions is informative, and $I$ can be estimated and reported alongside them.

%We use the first decomposition because its NDE is the direct effect in a world where treatment does not affect the competing event, the quantity of primary interest in our applications; $I$ can be estimated and reported alongside if desired.

\begin{remark}[Role of the no-unmeasured-confounding assumption]\label{remark: role of U natural effects}
Assumptions~\ref{ass:nat}(ii)--(iv) are needed to identify $F_{k,\nat}^{\overline{a}}(t)$ simultaneously for all $\overline{a} = (a_0, a_1, \ldots, a_J) \in \{0,1\}^{J+1}$, but a weaker subset of assumptions suffices for particular choices of $\overline{a}$. Table~S1 in the Supplementary Material A.4 summarizes which counterfactual CIFs and decomposed effects remain identified under specific violations of Assumptions~\ref{ass:nat}(ii)--(iv). For example, identification of $F_{k,\nat}^{\overline{a}}(t)$ for an $\overline{a}$ with $a_0 = a_j$ allows unmeasured common causes between the mediator and the type-$j$ first event, and identification for an $\overline{a}$ with $a_j = a_l$ allows unmeasured common causes between the type-$j$ and type-$l$ first events, for $j, l \in \{1, \ldots, J\}$.
\end{remark}

%As discussed in Supplementary Material A, we can identify another version of the direct effect in Setting 3 and another version of the competing effect in Setting 4.

%When $k$ varies from 1 to $J$, this decomposition gives the direct effect on each event and the effect through events other than this event.
%\textcolor{purple}{$\clubsuit$ PG 8-12-26: I think the above equations would be clearer if they include the info on what are the values of $k$ that these formulas can take. $\clubsuit$}
%For this three-component decomposition, $a_j$ is identical across all $j \neq k$, so the no-unmeasured-confounding assumption across competing events can be removed. 

%either by converting some would-be symptomatic infections into asymptomatic ones or via preventing symptomatic infections altogether, thereby altering the pools of individuals who remain at risk of asymptomatic infection.

\vspace{-0.7cm}
\subsection{Stochastic interventional effects} 
\label{sec:inteff}

%A limitation of the natural effects and separable effects frameworks is their inability to handle treatment-induced confounding and unobserved frailty between the mediator and survival events. When post-treatment confounding is present, the dismissible components condition generally requires partial isolation, saying that post-treatment confounding lies in isolated causal pathways from treatment components to their targets \citep{stensrud2021generalized}. In practice, the post-treatment confounding is typically not fully measured. Even if measured, the confounding may be affected by all treatment components, breaking down partial isolation. Separable effects are no longer identifiable from observed data. 

The stochastic interventional effects framework, like the natural effects framework, treats the treatment, mediator, and event-counting processes as manipulable. However, unlike natural effects, it avoids nested counterfactuals by generating the mediator and competing-event processes from researcher-specified external distributions.

%A generalization of the natural effects framework is the stochastic interventional effects framework. Interventional effects are based on the same philosophical stance as natural effects: the treatment, mediator, and event-counting processes are manipulable. Instead of assuming the mediator and counting processes of competing events are developed naturally under different treatment levels, the stochastic interventional effects framework replaces post-treatment variables with external versions. In a hypothetical world, we use $M^*$ and $dN_j^*(t)$ to denote the external mediator and event-$j$ counting process. A key feature of stochastic interventions is that we allow the mediator and event-counting processes to be generated according to arbitrary levels. While the ``natural'' levels of data generation are hard to define, stochastic interventions are always available with any data generation mechanism.

We use $M^*$ and $dN_j^*(t)$ to denote the mediator and the jump of the counting process for the type-$j$ first event, respectively. Specifically, $M^*$ is drawn from $f^*(m;x)$, a conditional density of $M$ given covariates $X=x$. Setting $f^*(m;x) = f(M^{a_0} \mid X=x)$ recovers the mediator as it would naturally arise under $A = a_0$. Alternatively, one may set $f^*(m;x)$ to a rightward shift of $f(M^{a_0} \mid X=x)$, reflecting a mediator generated as if under a possibly more potent treatment. On the other hand, for an individual at risk at time $t$, the conditional distribution of the increment $dN_j^*(t)$ is fully specified by some user-specified hazard function $\lambda_{j}^{*}(t;m,x)$, for each $j \neq k$:
\begin{equation*}
\lambda_{j}^{*}(t;m,x) := \lim_{dt\to0} \frac{1}{dt} P\left\{dN_j^{*}(t)=1 \Bigm| \sum_{l\neq k}N_l^{*}(t^-)=0, N_{k}^{a_k,m,\widetilde{N}_{-k}^*(t^-)}(t^-)=0, M^{*}=m, X=x\right\}.
\end{equation*}

By setting the treatment at $a_k \in \{0, 1\}$, replacing the mediator with $M^*$, and setting the counting processes of competing events to $\widetilde{N}_{-k}^*(t^-) = \{\{N_j^*(s): s<t\}, j = 1, \cdots, J, ~j \neq k\}$ via specifying hazard functions, we obtain $N_k^{a_k,M^*,\widetilde{N}_{-k}^*(t^-)}(t)$, the potential counting process for the type-$k$ first event. Our target estimand is the counterfactual type-$k$-specific CIF, $F_{k,\rnt}^{a_k*}(t) = P\{N_k^{a_k,M^*,\widetilde{N}_{-k}^*(t^-)}(t)=1\}$, evaluated in the counterfactual world in which treatment is set to $a_k$, the mediator is drawn from the density $f^*(m;x)$, and, for each competing event $j \neq k$, its counting process is drawn from the hazard function $\lambda_j^*(t;m,x)$. Let the total cumulative hazard of all competing events under intervention be $\Lambda_{-k}^*(t;m,x) = \sum_{j\neq k}\int_0^t\lambda_j^*(s;m,x)ds$. Theorem \ref{thm:int} identifies the target parameter $F_{k,\rnt}^{a_k*}(t)$.

%While the mediator and competing events $j \neq k$ are replaced with external counterparts, Assumptions \ref{ass:nat}(1) and \ref{ass:nat}(3) are still required to identify the type-$k$-specific hazard.

\begin{theorem} \label{thm:int}
Let $M^*$ be drawn from the density $f^*(m;x)$ and $dN_j^*(t)$ from the hazard $\lambda_j^*(t;m,x)$ for each $j\neq k$. Under Assumptions~\ref{ass:std} and \ref{ass:nat}(i), (ii), (iv), and (v), the type-$k$-specific counterfactual CIF $F_{k,\rnt}^{a_k*}(t) = P\{N_k^{a_k,M^*,\widetilde{N}_{-k}^*(t^-)}(t)=1\}$ is identified as follows:
\[
F_{k,\rnt}^{a_k*}(t) = \int_{\mathcal{X}} \int_{\mathcal{M}} \int_0^t \exp\left\{-\Lambda_k(s;a_k,m,x)-\Lambda_{-k}^*(s;m,x)\right\} d\Lambda_k(s;a_k,m,x) f^*(m;x) dm dP(x),
\]
where $\mathcal{X}$, $\mathcal{M}$, $\Lambda_k(s;a_k,m,x)$, and $P(x)$ are defined in Section \ref{sec:nateff}.
\end{theorem}

\begin{remark}[Identification assumptions for $F_{k,\rnt}^{a_k*}(t)$]
Identification under the stochastic interventional effects framework requires only a weaker, single-world version of Assumption~\ref{ass:nat}. Because the intervention draws the mediator $M^*$ from an externally specified distribution rather than setting it to its counterfactual value $M^{a_0}$, Theorem~\ref{thm:int} needs only the single-world components of Assumption~\ref{ass:nat}(i), (ii), and (iv) at treatment level $a_k$, together with no assumption linking $M^{a_0}$ to the potential event processes under a different treatment level; Assumption~\ref{ass:nat}(iii) is not needed at all. Moreover, because the intervention replaces the competing-event processes with stochastic processes generated from the externally specified hazards, and replaces the mediator with draws from the externally specified density, these intervened quantities do not depend on any unmeasured confounders. Consequently, unmeasured common causes of the mediator and the competing events, or of the type-$k$ first event and the instantaneous competing events, do not prevent identification.
\end{remark}

Different choices of $f^*(m;x)$ and $\lambda_j^*(t;m,x)$ answer different scientific questions. We recommend setting these to reflect data-generating processes that keep the resulting counterfactual world interpretable. Suppose we take $f^*(m;x) = f(M^{a_0};x)$, the density of $M$ under $A = a_0$ given $X = x$, and $\lambda_j^*(t;m,x) = \lambda_j(t;a_j,m,x)$, the type-$j$ hazard under $A = a_j$ given $M = m$ and $X = x$, and fix $A = a_k$. The counterfactual CIF $F_{k,\rnt}^{\overline{a}}(t) = P\{N_k^{a_k,M^*,\widetilde{N}_{-k}^*(t^-)}(t)=1\}$ then reduces to the same statistical functional as under the natural effects framework in Theorem~\ref{thm:nat}. As with natural effects, the total effect on the type-$k$-specific CIF decomposes into an interventional effect through the mediator (IME), an interventional direct effect (IDE), and an interventional effect through the competing events (ICE):
\begin{align*}
&\tau_k(t) := F_k^{a=1}(t) - F_k^{a=0}(t) = \IIM_k(t) + \IDE_k(t) + \IIC_k(t),\qquad\text{where} \\
\IIM_k(t) = &F_{k,\rnt}^{\bm{e}_0}(t) - F_{k,\rnt}^{\overline{0}}(t), \ 
\IDE_k(t) = F_{k,\rnt}^{\bm{e}_k}(t) - F_{k,\rnt}^{\bm{e}_0}(t), \ 
\IIC_k(t) = F_{k,\rnt}^{\overline{1}}(t) - F_{k,\rnt}^{\bm{e}_k}(t).
\end{align*}

%One technical nuance is worth mentioning: unlike the natural effects framework, which precludes unmeasured confounding, the distribution $f^*(m;x)$ and hazard functions $\lambda_j^*(t;m,x)$ can be interpreted as already marginalized over the unmeasured confounding. 

One can also anchor the specification in the identified conditional distribution of the observed  $M$ and the actual conditional hazard functions, and then perturb them to study other meaningful counterfactual worlds. For instance, in our running example of immune correlates analysis, if we set $f^*(m;x) = f(m-\delta;1,x)$ for some $\delta > 0$ and $\lambda_2^*(t;m,x) = 0.5\lambda_2(t;1,m,x)$, then this specification yields the CIF for asymptomatic infection in a counterfactual vaccinated world where vaccination elicits a stronger immune response (to the extent of a rightward shift by $\delta$ in the immune response distribution under current vaccination), the conditional hazard of symptomatic infection is halved, and the conditional hazard of asymptomatic infection remains unchanged. Another meaningful choice is the marginal hazard $\lambda_2^*(t; m, x) = \lambda_2^*(t; x) := \int_{\mathcal{M}} \lambda_2(t; 0, m, x)dP(m \mid T\geq t, A=0, X=x)$, obtained by marginalizing over the distribution of $M$ in the control group \citep{deng2026randomized}.

\vspace{-0.5cm}
\subsection{Separable effects} 
\label{sec:sepeff}
The separable effects framework conceptualizes that the treatment $A$ has $J+1$ components $A_0$, $A_1$, $\ldots$, $A_J$, where the component $A_0$ only directly influences the mediator, and the component $A_j$ only directly influences the counting process of event $j$ ($j=1,\ldots,J$). For example, if treatment $A$ represents chemotherapy for cancer treatment, $M$ represents severe toxicity, and the outcome of interest is overall survival, then a separable effects framework could view the chemotherapy as consisting of $2$ components, one that affects toxicity, and the other affecting the overall survival (e.g., via killing cancer cells). In a realized two-arm experiment, all treatment components are identical; that is, $A_0=\cdots=A_J = A$; however, in future studies with an updated treatment, these components can in principle be assigned different values. In the chemotherapy example, a future chemotherapy may have a modified toxicity-inducing component with less toxicity, and an improved cancer-cell-killing component.

Let $\overline{a} = (a_0, a_1, \ldots, a_J) \in \{0, 1\}^{J+1}$ denote the vector of treatment components. Let $M^{\overline{a}}$ denote the potential mediator under $\overline{a}$, and let $dN_j^{\overline{a}}(t)$ denote the increment over $[t, t + dt)$ of the potential counting process for the type-$j$ first event under $\overline{a}$, so that $N_j^{\overline{a}}(t) = \int_0^t dN_j^{\overline{a}}(s)$. As before, our interest lies in the counterfactual CIF of the type-$k$ first event under $\overline{a}$, $F_{k,\sep}^{\overline{a}}(t) = P\{N_k^{\overline{a}}(t) = 1\}$. A key identification assumption in the separable effects framework is the following dismissible components condition.

\begin{assumption} \label{ass:dcc}
(i) Ignorability of treatment assignment, $\{M^{\overline{a}}, N_j^{\overline{a}}(\cdot): j=1,\ldots,J\} \perp A \mid X$; (ii) dismissible components:
\begin{enumerate}
\item[(a)] for the mediator, $M^{(a_0,\ldots,a_J)} = M^{(a_0,\ldots,a_0)} = M^{a_0}$;
\item[(b)] for each event type $j=1,\ldots,J$, $N_j^{(a_j,\ldots,a_j)}(t) = N_j^{a_j}(t)$ and 
\begin{align*}
&\quad ~ \lim_{dt\to0} \frac{1}{dt} P\left\{dN_j^{(a_0,\ldots,a_J)}(t)=1 \bigm| \sum_{l=1}^{J}N_l^{(a_0,\ldots,a_J)}(t^-)=0, M^{(a_0,\ldots,a_J)}=m, X=x\right\} \\
&= \lim_{dt\to0} \frac{1}{dt} P\left\{dN_j^{(a_j,\ldots,a_j)}(t)=1 \bigm| \sum_{l=1}^{J}N_l^{(a_j,\ldots,a_j)}(t^-)=0, M^{(a_j,\ldots,a_j)}=m, X=x\right\}.
\end{align*}
\end{enumerate}
\end{assumption}

Assumption \ref{ass:dcc} states that the treatment components $(a_0,a_1,\ldots,a_J)$ have isolated effects on the mediator and hazards for each type of event \citep{stensrud2021generalized, stensrud2022separable, martinussen2021estimation}. Supplementary Figure~S2 illustrates the separable effects framework and the dismissible components condition using an extended DAG and the corresponding SWIG. The dismissible components condition may fail if there is treatment-induced confounding or if there are unmeasured common causes of the mediator and the event processes; nevertheless, separable effects remain well defined provided that the component interventions are well defined. The condition does allow for an unmeasured frailty that affects a single counting process across time points. As noted by \citet{stensrud2022separable}, although the dismissible components condition cannot be verified from the observed data, it could in principle be tested in a future trial in which the treatment components are manipulated independently.
%Figure~\ref{fig2}(a) illustrates the separable effects framework and the dismissible components assumption using an extended directed acyclic graph (eDAG) with three treatment components $A = (A_0, A_1, A_2)$ and two survival events $\{N_1(\cdot), N_2(\cdot)\}$. Figure~\ref{fig2}(b) further presents the single-world intervention graph (SWIG) to illustrate the relationship among potential outcomes. The treatment components associated with survival outcomes have no effects on $M$ by the dismissible components condition, which is plausible because the mediator is measured shortly after treatment. Conditional on the mediator, changing the treatment components other than $a_j$ does not alter the hazard of the type-$j$ first event; {\color{purple}the treatment component $A_j$ and the mediator $M$ together block} the path from $A$ to $N_j(t)$ when $N_1(t^-)=N_2(t^-)=0$.
%\begin{figure}
%    \centering
%    \includegraphics[width=0.9\textwidth]{fig2}
%    \caption{{\small Extended directed acyclic graph (a) and single-world intervention graph (b) under the separable effects framework.}} \label{fig2}
%\end{figure}
Theorem \ref{thm:sep} identifies $F_{k,\sep}^{\overline{a}}(t)$ under the separable effects framework. 

\begin{theorem} \label{thm:sep}
Under Assumptions \ref{ass:std} and \ref{ass:dcc}, the type-$k$-specific counterfactual CIF $F_{k,\sep}^{\overline{a}}(t) = P\{N_{k}^{\overline{a}}(t)=1\}$ is identified for $(a_0, \ldots, a_J) \in\{0,1\}^{J+1}$ as follows:
\[
F_{k,\sep}^{\overline{a}}(t) = \int_{\mathcal{X}} \int_{\mathcal{M}} \int_0^t \exp\left\{-\sum_{j=1}^{J}\Lambda_j(s;a_j,m,x)\right\} d\Lambda_k(s;a_k,m,x) dP(m;a_0,x) dP(x),
\]
where $\Lambda_j(t;a,m,x)$, $P(m;a,x)$, and $P(x)$ are observed-data functions defined in Section 2.2.
\end{theorem}

%Due to dismissible components, the mediator density is $f(m;a_0,x)$ and the event-$j$-specific hazard is $\Lambda_j(t;a_j,m,x)$ under the treatment components combination $\overline{a}$.

We note that the identification formula of $F_{k,\sep}^{\overline{a}}(t)$ under the separable effects framework is the same statistical functional as that in Theorem~\ref{thm:nat} under the natural effects framework, although the two frameworks differ in their conceptualization and identification assumptions. Moreover, as in the natural effects framework (see Remark~\ref{remark: role of U natural effects}), identification of $F_{k,\sep}^{\overline{a}}(t)$ for certain choices of $\overline{a}$ only requires a subset of Assumption~\ref{ass:dcc}; see Supplementary Material A.4 for details. 

When all components of $\overline{a}$ are identical, the counterfactual CIF reduces to $F_k^{(a,\ldots,a)}(t) = F_k^a(t)$, which corresponds to the observed type-$k$-specific CIF under treatment $A=a$. The total effect $\tau_k(t)$ can be decomposed into a separable effect through the mediator (SME), a separable direct effect (SDE), and a separable effect through competing events (SCE),
\begin{align*}
&\tau_j(t) := F_k^{a=1}(t) - F_k^{a=0}(t) 
= \SIM_k(t) + \SDE_k(t) + \SIC_k(t),\qquad\text{where}\\
\SIM_k(t) = &F_{k,\sep}^{\bm{e}_0}(t) - F_{k,\sep}^{\overline{0}}(t), \ 
\SDE_k(t) = F_{k,\sep}^{\bm{e}_k}(t) - F_{k,\sep}^{\bm{e}_0}(t), \ 
\SIC_k(t) = F_{k,\sep}^{\overline{1}}(t) - F_{k,\sep}^{\bm{e}_k}(t).
\end{align*}
The three terms $(\SIM_k, \SDE_k, \SIC_k)$ take the same form as $(\NIM_k, \NDE_k, \NIC_k)$ under the natural effects framework, but the interpretation of $(\SIM_k, \SDE_k, \SIC_k)$ is more straightforward. For instance, $\SIM_k$ represents the difference in the counterfactual CIF of the type-$k$ first event between a hypothetical treatment with components $(1, 0, \ldots, 0)$ and a hypothetical treatment with all-zero components. The other two terms admit analogous interpretations.

%\subsection{Comparing the three frameworks}
%\label{subsec: comparison}
Table~\ref{tab:compare1} compares the natural, stochastic interventional, and separable effects frameworks.

%While Assumptions \ref{ass:nat}(i), (ii) and (iii) are all concerned about the unmeasured confounding among the mediator, primary survival endpoint and the competing events. Let $k = 1$ denote the primary endpoint. 
%We need Assumptions \ref{ass:nat}(i), (ii) and (iii) to identify the counterfactual CIF $F_{1,\nat}^{\overline{a}}(t)$ for an arbitrary $\overline{a}$. When a subset of these assumptions fail, some counterfactual CIFs and decomposed effects remain identifiable. Table~\ref{tab:compare2} lists which counterfactual CIFs and decomposed treatment effects are identifiable when the no-unmeasured-confounding assumption fails.  %In the third setting where there is unmeasured confounding between the mediator and a competing event, the direct effect is identifiable if it is re-defined as $F_1^{(0,1,0,\ldots,0)}(t)-F_1^{\overline{0}}(t)$; in the fourth setting where there is unmeasured confounding between the mediator and primary endpoint, the competing effect is identifiable if it is re-defined as $F_1^{(0,0,1,\ldots,1)}(t)-F_1^{\overline{0}}(t)$.

%\textcolor{purple}{$\clubsuit$ PG 8-12-26: In Table 1, for the row `Key assumption for identifiability', I'm not sure why the positivity/overlap assumption is not considered just as key as sequential ignorability, given that the target parameter isn't defined unless positivity/overlap holds. $\clubsuit$}

% Preamble: \usepackage{booktabs, tabularx, array}
\newcolumntype{L}{>{\raggedright\arraybackslash}X}

\begin{table}[!tbh]
\centering
\small
\caption{Comparison of the natural, stochastic interventional, and separable effects frameworks.}
\label{tab:compare1}
\begin{tabularx}{\textwidth}{@{} >{\raggedright\arraybackslash}p{0.17\textwidth} L L L @{}}
\toprule
 & Natural effects & Stochastic interventional effects & Separable effects \\
\midrule
What is intervened on
 & Treatment, mediator, and event processes
 & Treatment, mediator, and event processes
 & Treatment components \\
%\addlinespace
How the intervention works
 & Mediator and event processes are set to values generated under (possibly different) treatment levels
 & Mediator and event processes are drawn from specified distributions
 & Treatment components are assigned different levels \\
%\addlinespace
Common identification assumptions
 & Assumption~\ref{ass:std}
 & Assumption~\ref{ass:std}
 & Assumption~\ref{ass:std} \\
%\addlinespace
Additional identification assumption
 & Sequential ignorability (cross-world)
 & Sequential ignorability (weaker, single-world version)
 & Dismissible components condition \\
%\addlinespace
Implication of the additional assumption
 & No unmeasured confounding between (i) the mediator and any event process or (ii) any two event processes
 & No unmeasured confounding between (i) the mediator and the primary event or (ii) the primary event and any competing event
 & No unmeasured confounding between (i) the mediator and any event process or (ii) any two event processes \\
%\addlinespace
Testability of the additional assumption
 & Not testable
 & Not testable
 & Testable in principle in future trials \\
%\addlinespace
% Identification formula & Equation~\eqref{eq: survival g-formula} & ... \\
Limitations
 & Requires cross-world assumptions
 & Effect may not be sharp under treatment-induced confounding, even if identifiable
 & Requires a treatment decomposable into distinct components %which may not exist in practice 
 \\
\bottomrule
\end{tabularx}
\end{table}

\vspace{-0.7cm}
\section{Controlled effects} 
\label{sec:cnteff}
In addition to effect decomposition, a researcher may also be interested in the dose--response relationship between the mediator and the counterfactual risk. For example, in an immune correlates analysis, \citet{gilbert2023controlled} studied how the counterfactual risk varies under a hypothetical joint intervention that vaccinates the population ($A = 1$) and sets the vaccine-induced immune marker to a given level ($M = m$). In a competing-risks setting, at least two controlled-risk estimands may be of interest. First, researchers may want to estimate the counterfactual CIF of the type-$k$ first event under a hypothetical joint intervention that assigns treatment $A = a$ to the population and sets the mediator at a user-specified level ($M = m_0$), while allowing the primary and competing event processes to evolve according to their counterfactual laws under the joint intervention $(A=a, M=m_0)$. This scenario can be viewed as a hybrid of the interventional effects and the natural effects frameworks, with the mediator distribution set to a degenerate distribution at the singleton $\{m_0\}$, and the event hazard functions left at natural levels under $A = a$. The resulting controlled counterfactual type-$k$-specific CIF is identified under Assumptions \ref{ass:std} and \ref{ass:nat} as follows:
\[
F_k^{a,m_0}(t) = \int_{\mathcal{X}}\int_0^t \exp\{-\Lambda(s;a,m_0,x)\} d\Lambda_k(s;a,m_0,x)dP(x),
\]
where $\Lambda(t;a,m,x) = \sum_{j=1}^{J}\int_0^t\lambda_j(s;a,m,x)ds$ is the all-cause cumulative hazard.

%is the cumulative hazard of the first event of any type.
%Estimating the controlled counterfactual risk is especially relevant to a key application in immunobridging research, where researchers seek to understand how effective a future vaccine would be if it could elicit a more potent immune response, provided that there is no relative direct effect between the investigated and future vaccines. 
%\textcolor{purple}{PG 8-12- 26: On `this approach is especially relevant,' not necessarily, see my first comment.  Also, the controlled risk parameter of Gilbert et al. (2023) is essentially the same as controlled effects as introduced long ago (like Robins and Greenland, 1992), so wouldn't want to convey a tone that Gilbert et al. (2023) did something generally new in introducing this controlled effects parameter.}

%Existing approaches to address inference on this controlled quantity include model-based inference, which may lose flexibility, and kernel smoothing \citep{kennedy2017non, kenny2025inference}. 

As discussed in Supplementary Material D, $F_k^{a,m_0}(t)$ is not pathwise differentiable in a nonparametric model when the mediator is continuous. Instead of directly targeting $F_k^{a,m_0}(t)$, we consider the following kernel-smoothed controlled counterfactual type-$k$-specific CIF (see, e.g., \citet{branson2023causal}):
\begin{align*}
F_{k,K_h}^{a,m_0}(t) = \int_{\mathcal{M}} K_h(m-m_0) F_k^{a,m}(t) dm,
\end{align*}
where $K_h(u)=K(u/h)/h$ is a symmetric kernel function with bandwidth $h$, subject to $\int_{-\infty}^{\infty}K_h(u)du=1$ and $\int_{-\infty}^{\infty}K_h(u)^2du<\infty$. This kernel-smoothed CIF can also be interpreted as the CIF under the stochastic interventional framework, with the mediator drawn from the density $f^*(m;x) = K_h(m-m_0)$ and hazards of competing events left at their natural values.

Alternatively, researchers may study controlled effects that additionally set the type-specific hazards of competing events to $\{\lambda_{j}^*(t;m,x): j=1,\ldots,J,~j\neq k\}$, so as to examine the relationship between the mediator and the type-$k$-specific risk under counterfactual data-generating processes for the competing events. Let $F_k^{a,m_0,\Lambda_{-k}^*}(t)$ denote the controlled counterfactual type-$k$-specific CIF setting $A = a$, $M = m_0$, and the total cumulative hazard of all competing events $\Lambda_{-k}^*(t;m,x)$. Directly following Theorem~\ref{thm:int}, $F_k^{a,m_0,\Lambda_{-k}^*}(t)$ is identified as %follows:
\[
F_k^{a,m_0,\Lambda_{-k}^*}(t) = \int_{\mathcal{X}}\int_0^t \exp\{-\Lambda_{-k}^*(s;m_0,x)-\Lambda_k(s;a,m_0,x)\} d\Lambda_k(s;a,m_0,x)dP(x),
\]
and its kernel-smoothed version is $F_{k,K_h}^{a,m_0,\Lambda_{-k}^*}(t) = \int_{\mathcal{M}} K_h(m-m_0) F_k^{a,m,\Lambda_{-k}^*}(t) dm$, which is also the CIF under the stochastic interventional framework, with the mediator drawn from the density $f^*(m;x) = K_h(m-m_0)$ and competing events from cumulative hazards $\Lambda_{-k}^*(t;m,x)$.

\section{Efficient estimation and inference}
\label{sec: estimation and inference}
\vspace{-0.3cm}
\subsection{Counterfactual CIFs and decomposed effects}
\label{subsec: estimate decompoed effects}
Counterfactual CIFs $F_{k, \nat}^{\overline{a}}(t)$ and $F_{k, \sep}^{\overline{a}}(t)$ coincide in their expression, and they both coincide with with $F_{k,\rnt}^{a_k*}(t)$ when $f^*(m;x) = f(m;a_0,x)$ and $\lambda_j^*(t;m,x) = \lambda_j(t;a_j,m,x)$ in $F_{k,\rnt}^{a_k*}(t)$. Below, we denote $F_k^{\overline{a}}(t)$ as this counterfactual CIF and study its efficient estimation and inference. 
%To estimate $F_k^{\overline{a}}(t)$, the hazard functions and the mediator density should be estimated since they are unknown.

Denote the propensity score $\pi(a;x) = P(A=a \mid X=x)$, the cumulative censoring hazard $\Lambda_0(t;a,m,x)$, and the counterfactual cumulative hazard $\Lambda(t;\overline{a},m,x) = \sum_{j=1}^{J}\Lambda_j(t;a_j,m,x)$ for $\overline{a}=(a_0,a_1,\ldots,a_J)$. Define
\[
G_j(t) = \int_0^t \frac{dN_j(s) - Y(s) d\Lambda_j(s;A,M,X)}{P(T \geq s \mid A,M,X)},
\]
the integrated martingale residual associated with the type-$j$ first event inverse-weighted by the observed survival function. The denominator $P(T \geq s \mid A,M,X) = \exp\{-\Lambda(s^-;A,M,X)-\Lambda_0(s^-;A,M,X)\}$.
Lemma~\ref{lemma: EIF F_k^a} derives the efficient influence function (EIF) of $F_k^{\overline{a}}(t)$.

\begin{lemma}
\label{lemma: EIF F_k^a}
The efficient influence function of $F_k^{\overline{a}}(t)$ is
\begin{align*}
\varphi_k^{\overline{a}}(t;O) &= \frac{I(A=a_k)}{\pi(a_k;X)} \frac{f(M;a_0,X)}{f(M;a_k,X)} \int_0^t \exp\{-\Lambda(s;\overline{a},M,X)\}dG_k(s) \\
&\quad - \sum_{j=1}^{J} \frac{I(A=a_j)}{\pi(a_j;X)} \frac{f(M;a_0,X)}{f(M;a_j,X)} \int_0^t \exp\{-\Lambda(s;\overline{a},M,X)\} G_j(s) d\Lambda_k(s;a_k,M,X) \\
&\quad + \frac{I(A=a_0)}{\pi(a_0;X)} \int_0^t \exp\{-\Lambda(s;\overline{a},M,X)\} d\Lambda_k(s;a_k,M,X) \\
&\quad + \left\{1-\frac{I(A=a_0)}{\pi(a_0;X)}\right\} \int_0^t \int_{\mathcal{M}} \exp\{-\Lambda(s;\overline{a},m,X)\} d\Lambda_k(s;a_k,m,X) dP(m;a_0,X)  - F_k^{\overline{a}}(t).
\end{align*}
\end{lemma}

We denote the uncentered EIF as $\phi_k^{\overline{a}}(t;O)$, so that $\varphi_k^{\overline{a}}(t;O) = \phi_k^{\overline{a}}(t;O) - F_k^{\overline{a}}(t)$. The EIF involves four nuisance functions: the propensity score $\pi(a;x)$, the mediator density ratio $r(m,x,a_0,a_j)=f(m;a_0,x)/f(m;a_j,x)$, the cumulative type-specific hazards $\{\Lambda_j(t;a,m,x): j=1,\ldots,J\}$, and the cumulative censoring hazard $\Lambda_0(t;a,m,x)$. Evaluating the EIF requires fitting these models. When the mediator $M$ is continuous, the mediator density ratio $r(m,x,a_0,a_j)$ is usually unstable to estimate. We propose to estimate the density ratio as follows. By Assumption~\ref{ass:std}(iii), $r(M,X,a_0,a_j)$ is bounded. For any $L_2(\mathbb{P})$-integrable function $\gamma(M,X)$, we notice that
\[
\mathcal{F}(\gamma) := E\left[\frac{I(A=a_j)}{\pi(a_j;X)}r(M,X,a_0,a_j)\gamma(M,X)\right] = E\left[\frac{I(A=a_0)}{\pi(a_0;X)}\gamma(M,X)\right]
\]
is a linear operator for $\gamma$, so $r(M,X,a_0,a_j)$ is the Riesz representer for this operator under the weighted inner product defined on the left-hand side. To estimate the true $r_0(M,X,a_0,a_j) \in L_2(\mathbb{P})$, we approximate it in a function class $\mathcal{H}$ by solving
\begin{align*}
\widehat{r} &= \arg\min_{r\in\mathcal{H}} E\left[\frac{I(A=a_j)}{\pi(a_j;X)}\{r(M,X,a_0,a_j) - r_0(M,X,a_0,a_j)\}^2\right] \\
&= \arg\min_{r\in\mathcal{H}} E\left[\frac{I(A=a_j)}{\pi(a_j;X)}r(M,X,a_0,a_j)^2\right] - 2E\left[\frac{I(A=a_j)}{\pi(a_j;X)}r(M,X,a_0,a_j)r_0(M,X,a_0,a_j)\right] \\
&= \arg\min_{r\in\mathcal{H}} E\left[\frac{I(A=a_j)}{\pi(a_j;X)}r(M,X,a_0,a_j)^2\right] - 2E\left[\frac{I(A=a_0)}{\pi(a_0;X)}r(M,X,a_0,a_j)\right],
\end{align*}
which reduces to the unconstrained least-squares importance fitting if $\mathcal{H}$ is a basis function class \citep{kanamori2009least, sugiyama2012density, chernozhukov2021automatic}.
%In practice, it is easier to fit the propensity score model $\pi(a;m,x) = P(A=a \mid M = m, X = x)$, using $(M,X)$ as predictors and $A$ as the binary response. Evaluating the EIF also requires integrating over the domain of $M$, which can be approximated numerically.

%(e.g., $\textsf{SuperLearner}$ by \citet{Polley2024superlearner} and $\textsf{survSuperLearner}$ by \citet{survSuperLearner})

We use cross-fitting to estimate $F_k^{\overline{a}}(t)$ \citep{chernozhukov2018double}. We randomly divide the sample into two parts, $\mathcal{I}_1$ and $\mathcal{I}_2$, of approximately equal size. Using the data in $\mathcal{I}_l$ ($l = 1, 2$), we fit the propensity score $\pi(a; x)$ and the type-specific and censoring hazards $\Lambda_j^{(l)}(t; a, m, x)$ using ensemble machine learning methods, and fit the mediator density ratio $r(m, x, a_0, a_j)$ using a Riesz learner. We denote these estimators by $\widehat{\pi}^{(l)}(a; x)$, $\widehat\Lambda_j^{(l)}(t; a, m, x)$, and $\widehat{r}^{(l)}(m, x, a_0, a_j)$, respectively. Next, in the subsample $\mathcal{I}_{3-l}$, we evaluate the EIF by plugging in the fitted models and integrating over $M$ using the predicted conditional expectation. Let $\widehat\phi_k^{\overline{a}}(t;O_i) = \widehat\phi_k^{\overline{a},(l)}(t;O_i)$ be the fitted $\phi_k^{\overline{a}}(t;O_i)$ for $i\in\mathcal{I}_{3-l}$. An estimator of $F_k^{\overline{a}}(t)$ is then given by
\begin{equation*}
\widehat{F}_k^{\overline{a}}(t) = n^{-1} \sum_{i=1}^{n} \widehat\phi_k^{\overline{a}}(t;O_i) = n^{-1} \sum_{l\in\{1,2\}} \sum_{i\in\mathcal{I}_{3-l}} \widehat\phi_k^{\overline{a},(l)}(t;O_i).
\end{equation*}
The resulting estimator $\widehat{F}_k^{\overline{a}}(t)$ is multiply robust, as formalized by Theorem~\ref{thm:mr}.

\begin{theorem}[Multiple robustness] \label{thm:mr}
The estimator $\widehat{F}_k^{\overline{a}}(t)$ converges to $F_k^{\overline{a}}(t)$ uniformly in $t\in(0,t^*)$ if any of the following conditions holds:
\begin{enumerate}{\small
    \item[(i)] the estimators of $\{r(m,x,a_0,a_j), \Lambda_j(t;a,m,x): j=1,\ldots,J\}$ are uniformly $L_1$ consistent; 
    %i.e., the propensity score and censoring hazard models can be misspecified while other models are correctly specified.
    \item[(ii)] the estimators of $\{\pi(a;x), \Lambda_j(t;a,m,x):j=1,\ldots,J\}$ are uniformly $L_1$ consistent; 
    %i.e., the mediator density ratio and censoring hazard models can be misspecified while other models are correctly specified.
    \item[(iii)]for some $l \in \{1, \ldots, J\}$, the estimators of $\pi(a; x)$ and of $\{r(m, x, a_0, a_j), \Lambda_j(t; a, m, x) : j \in \{0, 1, \ldots, J\} \setminus \{l\}\}$ are uniformly $L_1$ consistent. }%i.e., one type-specific hazard model is misspecified while other models are correctly specified.
\end{enumerate}
In each case, the remaining nuisance estimators need not be consistent, provided that they converge uniformly to fixed limits.
\end{theorem}

%The proof is provided in Supplementary Material B. 

When all nuisance estimators are $L_2$ consistent at sufficiently fast rates, we have the following asymptotic property.

%The multiple robustness of the EIF-based estimator $\widehat{F}_k^{\overline{a}}(t)$ indicates that one of the working models can be misspecified while others are correctly specified.

\begin{theorem}[Asymptotic normality] \label{thm:asym}
Suppose that all nuisance estimators converge in $L_2$ at rate $o_P(n^{-1/4})$. Then $\sqrt{n}\left\{\widehat{F}_k^{\overline{a}}(\cdot) - F_k^{\overline{a}}(\cdot)\right\}$ converges weakly to a mean-zero Gaussian process in $\ell^\infty(0, t^*)$. In particular, for each $t \in (0, t^\ast)$,
\[
\sqrt{n} \left\{\widehat{F}_k^{\overline{a}}(t) - F_k^{\overline{a}}(t)\right\} \xrightarrow{d} N\left(0, E\{\varphi_k^{\overline{a}}(t;O)^2\}\right).
\]
The asymptotic variance attains the nonparametric efficiency bound.
\end{theorem}

%The proof is provided in Supplementary Material C. 
A wide range of semiparametric estimators can satisfy the convergence rate condition, including the maximum partial likelihood estimator for the Cox model with Breslow hazard estimator, sieve estimators under sufficient sparsity for mediator density ratio \citep{sugiyama2012density}, and random (survival) forests under sufficient smoothness \citep{wager2018estimation}. The standard error can be calculated based on the fitted efficient influence function,
\[
\mathrm{se}\left\{\widehat{F}_k^{\overline{a}}(t)\right\} = \frac{1}{\sqrt{n}} \left[\frac{1}{n-1}\sum_{i=1}^{n}\{\widehat\phi_k^{\overline{a}}(t;O_i)-\widehat{F}_k^{\overline{a}}(t)\}^2\right]^{1/2}.
\]

The decomposed effects --- $(\NIM_k, \NDE_k, \NIC_k)$ in Section \ref{sec:nateff}, $(\IIM_k, \IDE_k, \IIC_k)$ in Section \ref{sec:inteff}, and $(\SIM_k, \SDE_k, \SIC_k)$ in Section \ref{sec:sepeff} --- can then be estimated as contrasts of the corresponding $\widehat{F}_k^{\overline{a}}(t)$ under the appropriate choices of $\overline{a}$. The standard error of the decomposed effects can be calculated similarly from the contrast of EIFs. For inference, the $P$-values for testing whether the decomposed effects are null can be calculated from the restricted mean survival time lost at the end of the study; see Supplementary Material C for details.

Finally, Lemma \ref{lemma: EIF F_k_int} derives the EIF for the type-$k$-specific CIF with researcher-specified, known interventional distribution $f^*(m;x)$ and hazards $\{\lambda_j^*(t;m,x): j=1,\ldots,J,~j \neq k\}$. 

\begin{lemma}
\label{lemma: EIF F_k_int}
The efficient influence function of $F_{k,\text{int}}^{a_k*}(t)$ with  known $f^*(m;x)$ and $\Lambda_{-k}^*(t;m,x)$ is
{\small 
\begin{align*}
\varphi_{k,\text{int}}^{a_k*}(t;O) &= \frac{I(A=a_k)}{\pi(a_k;X)} \frac{f^*(M;X)}{f(M;a_k,X)} \int_0^t \exp\{-\Lambda_k(s;A,M,X)-\Lambda_{-k}^*(s;M,X)\}dG_k(s) \\
&\quad - \frac{I(A=a_k)}{\pi(a_k;X)} \frac{f^*(M;X)}{f(M;a_k,X)} \int_0^t \exp\{-\Lambda_k(s;A,M,X)
%&\qquad\qquad\qquad\qquad\qquad\qquad\qquad 
- \Lambda_{-k}^*(s;M,X)\} G_k(s) d\Lambda_k(s;A,M,X) \\
&+ \int_0^t \int_{\mathcal{M}} \exp\{-\Lambda_k(s;a_k,m,X)-\Lambda_{-k}^*(s;m,X)\} f^*(m;x) d\Lambda_k(s;a_k,m,X) dm - F_{k,\text{int}}^{a_k*}(t).
\end{align*}
}
\end{lemma}

Similarly, $F_{k,\text{int}}^{a_k*}(t)$ can be estimated via cross-fitting, denoted by $\widehat{F}_{k,\text{int}}^{a_k*}(t)$, which solves the estimating equation $\mathbb{P}_n{\varphi_{k,\text{int}}^{a_k*}(t;O)}=0$. The number of working models is not reduced even when $f^*(m;x)$ and $\Lambda^*_{-k}(t;m,x)$ are known, as $G_k(t)$ involves all-cause hazards, while $f(m;a,x)$ enters directly into the first two terms of the EIF. Under the conditions of Theorem \ref{thm:mr}, the EIF-based estimator $\widehat{F}_{k,\text{int}}^{a_k*}(t)$ is uniformly consistent over $t\in(0,t^\ast)$. Under the conditions of Theorem \ref{thm:asym}, $\widehat{F}_{k,\text{int}}^{a_k}(t)$ is asymptotically normal and nonparametrically efficient.

\subsection{Smoothed controlled effects}

%The EIFs of the controlled type-$k$-specific counterfactual CIFs $F_k^{a,m_0,\Lambda_{-k}^*}(t)$ and $F_k^{a,m_0}(t)$ involve an indicator function $I(M=m_0)$, which are not pathwise differentiable if $M$ is continuous; see Supplementary Material D. Instead, we consider kernel-smoothed controlled effects
%\begin{align*}
%F_{k,K_h}^{a,m_0,\Lambda_{-k}^*}(t) = \int_{\mathcal{M}} K_h(m-m_0) F_k^{a,m,\Lambda_{-k}^*}(t) dm, \
%F_{k,K_h}^{a,m_0}(t) = \int_{\mathcal{M}} K_h(m-m_0) F_k^{a,m}(t) dm,
%\end{align*}
%where $K_h(u)=K(u/h)/h$ is a symmetric kernel function with bandwidth $h$, subject to $\int_{-\infty}^{\infty}K_h(u)du=1$ and $\int_{-\infty}^{\infty}K_h(u)^2du<\infty$ \citep{branson2023causal}. 

%Write
%\begin{align*}
%F_k(t;a,m,\Lambda_{-k}^*,x) &:= \int_0^t \exp\{-\Lambda_{-k}^*(s;m,x)-\Lambda_k(s;a,m,x)\} d\Lambda_k(s;a,m,x), \\
%F_k(t;a,m,x) &:= \int_0^t \exp\{-\Lambda(s;a,m,x)\} d\Lambda_k(s;a,m,x).
%\end{align*}
%as $\widehat{F}_k(t;a,m,\Lambda_{-k}^*,x)$ and $\widehat{F}_k(t;a,m,x)$, respectively. 
Let $F_k(t;a,m,\Lambda_{-k}^*,x) = \int_0^t \exp\{-\Lambda_{-k}^*(s;m,x)-\Lambda_k(s;a,m,x)\} d\Lambda_k(s;a,m,x)$ and $F_k(t;a,m,x) = \int_0^t \exp\{-\Lambda(s;a,m,x)\} d\Lambda_k(s;a,m,x)$. Lemma~\ref{lemma: EIF smoothed control effect} gives the EIFs of kernel-smoothed controlled effect parameters.

\begin{lemma}
\label{lemma: EIF smoothed control effect}
The efficient influence functions of $F_{k,K_h}^{a,m_0,\Lambda_{-k}^*}(t)$ and $F_{k,K_h}^{a,m_0}(t)$ are
{\small
\begin{align*}
\varphi_{k,K_h}^{a,m_0,\Lambda_{-k}^*}(t) &= \frac{I(A=a)}{\pi(a;X)}\frac{K_h(M-m_0)}{f(M;a,X)} \int_0^t \exp\{-\Lambda_{-k}^*(s;M,X)-\Lambda_k(s;A,M,X)\} dG_k(s) \\
&\quad - \frac{I(A=a)}{\pi(a;X)}\frac{K_h(M-m_0)}{f(M;a,X)} \int_0^t \exp\{-\Lambda_{-k}^*(s;M,X)-\Lambda_k(s;A,M,X)\} G_k(s) d\Lambda_k(s;A,M,X) \\
&\quad + \int_{\mathcal{M}}K_h(m-m_0) F_k(t;a,m,\Lambda_{-k}^*,X)dm - F_{k,K_h}^{a,m_0,\Lambda_{-k}^*}(t), \\
\varphi_{k,K_h}^{a,m_0}(t) &= \frac{I(A=a)}{\pi(a;X)}\frac{K_h(M-m_0)}{f(M;a,X)} \int_0^t \exp\{-\Lambda(s;A,M,X)\} dG_k(s) \\
&\quad - \frac{I(A=a)}{\pi(a;X)}\frac{K_h(M-m_0)}{f(M;a,X)} \sum_{j=1}^{J} \int_0^t \exp\{-\Lambda(s;A,M,X)\} G_j(s) d\Lambda_k(s;A,M,X) \\
&\quad + \int_{\mathcal{M}}K_h(m-m_0) F_k(t;a,m,X)dm - F_{k,K_h}^{a,m_0}(t).
\end{align*}
}
\end{lemma}

The kernel-smoothed controlled CIFs can be estimated by solving the estimating equations given by the EIFs, with the fitted nuisance models plugged in. Theorem \ref{thm:controlled CIFs} establishes the asymptotic properties of the resulting estimators $\widehat{F}_{k,K_h}^{a,m_0,\Lambda_{-k}^*}(t)$ and $\widehat{F}_{k,K_h}^{a,m_0}(t)$.

%are root-$n$ consistent for $F_{k,K_h}^{a,m_0,\Lambda_{-k}^*}(t)$ and $F_{k,K_h}^{a,m_0}(t)$, respectively. Their asymptotic properties are established as follows.

\begin{theorem}
\label{thm:controlled CIFs}
Suppose all nuisance estimators converge in $L_2$ at rate $o_p(n^{-1/4})$. Let
\[
\mathcal{M}_{K_h} = \left\{m_0:\ \frac{K_h(M-m_0)}{\widehat{f}(M;a,X)} > c
\ \text{ and }\ \frac{K_h(M-m_0)}{f(M;a,X)} > c \ \text{ almost surely}\right\},
\]
where $c>0$ is a constant. Then, for any $m_0 \in \mathcal{M}_{K_h}$, the processes $\sqrt{n}\left\{\widehat{F}_{k,K_h}^{a,m_0,\Lambda_{-k}^*}(\cdot) - F_{k,K_h}^{a,m_0,\Lambda_{-k}^*}(\cdot)\right\}$ and $\sqrt{n}\left\{\widehat{F}_{k,K_h}^{a,m_0}(\cdot) - F_{k,K_h}^{a,m_0}(\cdot)\right\}$ converge weakly to mean-zero Gaussian processes in $\ell^\infty(0, t^*)$. In particular, for each $t \in [0, t^*]$,
\begin{align*}
\sqrt{n}\left\{\widehat{F}_{k,K_h}^{a,m_0,\Lambda_{-k}^*}(t) - F_{k,K_h}^{a,m_0,\Lambda_{-k}^*}(t)\right\} &\xrightarrow{d} N\left(0, E\left[\left\{\varphi_{k,K_h}^{a,m_0,\Lambda_{-k}^*}(t)\right\}^2\right]\right), \\
\sqrt{n}\left\{\widehat{F}_{k,K_h}^{a,m_0}(t) - F_{k,K_h}^{a,m_0}(t)\right\} &\xrightarrow{d} N\left(0, E\left[\left\{\varphi_{k,K_h}^{a,m_0}(t)\right\}^2\right]\right).
\end{align*}

%Then, for any $m_0 \in \mathcal{M}_{K_h}$, the estimators $\widehat{F}_{k,K_h}^{a,m_0,\Lambda_{-k}^*}(\cdot)$ and $\widehat{F}_{k,K_h}^{a,m_0}(\cdot)$ converge weakly to Gaussian processes on $t \in (0,t^*)$, with
%\begin{align*}
%\sqrt{n}\left\{\widehat{F}_{k,K_h}^{a,m_0,\Lambda_{-k}^*}(t) - F_{k,K_h}^{a,m_0,\Lambda_{-k}^*}(t)\right\} &\xrightarrow{d} N\left(0, E\{\varphi_{k,K_h}^{a,m_0,\Lambda_{-k}^*}(t)^2\}\right), \\
%\sqrt{n}\left\{\widehat{F}_{k,K_h}^{a,m_0}(t) - F_{k,K_h}^{a,m_0}(t)\right\} &\xrightarrow{d} N\left(0, E\{\varphi_{k,K_h}^{a,m_0}(t)^2\}\right).
%\end{align*}
\end{theorem}

\vspace{-1cm}
\section{Extensions}
\label{sec: extension}
\vspace{-0.3cm}
\subsection{Two-stage sampling of the mediator}
\label{subsec: two stage sampling of M}
Measuring mediators --- for example, immune responses in vaccine trials --- can be expensive and cumbersome; hence, the mediator is often measured in only a subset of participants, selected under a prespecified, possibly outcome-dependent sampling scheme. In our running example, we collect peripheral blood mononuclear cell (PBMC) or serum samples after vaccination and store them, then run assays at the end of the study on sampled participants under a stratified case-cohort or case-control sampling scheme. In Supplementary Material E.1, we show that the EIFs of $F_k^{\overline{a}}(t)$, $F_{k,K_h}^{a,m_0,\Lambda_{-k}^*}(t)$, and $F_{k,K_h}^{a,m_0}(t)$ can be modified to account for two-stage sampling in a unified way \citep{rose2011targeted}. 

\subsection{Sensitivity analysis}
\label{subsec: SA}
As discussed in Remark~\ref{remark: role of U natural effects} and Supplementary Material~A.4, the sum of NDE and NCE remains identifiable under unmeasured confounding between the primary endpoint and the competing events; Assumptions~\ref{ass:nat}(iii)--(iv) are what allow this sum to be further decomposed into NDE and NCE. We therefore propose a sensitivity analysis that relaxes Assumptions~\ref{ass:nat}(iii)--(iv) by permitting unmeasured common causes of the primary endpoint and the competing events, and examine how the decomposition varies with the sensitivity parameter. Plausible unmeasured common causes between primary and competing events fall into three main categories: (a) intensity of exposure to SARS-CoV-2; (b) history of immunity-conferring events, such as prior SARS-CoV-2 infections and COVID-19 vaccinations; and (c) other baseline immunological factors, such as host genetic or microbiome-related susceptibility. %{\color{blue}Two major categories of plausible unmeasured common causes are (a) a history of immunity-conferring events, such as prior SARS-CoV-2 infections and COVID-19 vaccinations; and (b) other baseline immunological factors, such as host immune genetics, transcriptomics, or microbiome-related susceptibility.}

We posit an unmeasured common cause $\Xi \sim N(0, 1)$ that influences the hazard of all survival events. We assume that $\Xi$ is independent of $(A, M, X)$, and that Assumptions~\ref{ass:nat}(iii)--(iv) hold further conditional on $\Xi$. We posit the following model 
\begin{equation} 
\label{eq:hazard_sensitivity}
\begin{split}
\lambda_j(t;a,m,x,\xi) := &\lim_{dt\to0} P(t\leq T<t+dt, \Delta=j \mid T\geq t, A=a, M=m, X=x, \Xi = \xi)/dt \\
= &\exp(\gamma_j\xi) \lambda_j^0(t;a,m,x),
\end{split}
\end{equation}
where $\gamma_j$ captures the effect of $\Xi$ on the type-$j$-specific hazard, for $j=1,\ldots,J$. The posited latent variable $\Xi$ captures the latent correlation across all survival events unexplained by observed variables. Given $\Gamma = \{\gamma_j: j=1,\ldots,J\}$, the type-$k$-specific counterfactual CIF is
\begin{equation}
\label{eq: sensitivity analysis}
\begin{aligned}
    F_{k,\nat}^{\overline{a}}(t;\Gamma) &= \int_{\xi} \int_{\mathcal{X}} \int_{\mathcal{M}} \int_0^t \frac{1}{(2\pi)^{1/2}} \exp\left(-\frac{\xi^2}{2}\right) \exp\left\{\gamma_k\xi-\sum_{j=1}^{J}e^{\gamma_j\xi}\Lambda_j^0(s;a_j,m,x)\right\} \\
    &\qquad\qquad\qquad \times d\Lambda_k^0(s;a_k,m,x) dP(m;a_0,x) dP(x) d\xi.
\end{aligned}
\end{equation}

For any fixed $\Gamma$, the observed-data model induced by \eqref{eq:hazard_sensitivity} with $\{\lambda_j^0: j=1,\ldots,J\}$ unrestricted coincides with the nonparametric model, so $\Gamma$ is not identified; hence, it is appropriate to treat $\Gamma$ as a sensitivity parameter. When $a_1=\cdots=a_J=a$, integrating \eqref{eq: sensitivity analysis} over $\xi$ yields $F_{k,\nat}^{(a_0,a,\ldots,a)}(t;\Gamma) = F_{k,\nat}^{(a_0,a,\ldots,a)}(t)$; in particular, NME and NDE + NCE remain unchanged over $\Gamma$. The magnitude of $\Gamma$ determines how NDE + NCE is allocated between the NDE and NCE components. If we choose $\gamma_j\equiv\gamma$ for each $j$, then the $\gamma$ that makes the effects insignificant at a prespecified level serves as a tipping point for the robustness to unmeasured frailty. %The natural direct effect is $\NDE_k(t;\Gamma) = F_{k,\nat}^{\bm{e}_k}(t;\Gamma) - F_{k,\nat}^{\bm{e}_0}(t;\Gamma)$, and the natural competing effect is $\NIC_k(t;\Gamma) = F_{k,\nat}^{\overline{1}}(t;\Gamma) - F_{k,\nat}^{\bm{e}_k}(t;\Gamma)$.
Analogously, in the separable effects framework, the sum $\mathrm{SDE} + \mathrm{SCE}$ remains identified when Assumption~\ref{ass:dcc}(ii)(b) fails. The procedure above then also serves as a sensitivity analysis for Assumption~\ref{ass:dcc}(ii)(b): varying the sensitivity parameter $\Gamma$ yields a decomposition of $\mathrm{SDE} + \mathrm{SCE}$ into its two components for each value of $\Gamma$.

By varying $\Gamma$, the $\Xi$-independent part $\{\lambda_j^0: j=1,\ldots,J\}$ can be estimated under the model constraint \eqref{eq:hazard_sensitivity}. Let $\widehat\phi_k^{\overline{a}}(t;O_i,\xi,\Gamma)$ denote the uncentered EIF, in which $\widehat\Lambda_j(t;a,m,x)$ is replaced with $\exp(\gamma_j\xi)\widehat\Lambda_j^0(t;a,m,x)$ in $\widehat\phi_k^{\overline{a}}(t;O_i)$. The counterfactual CIF in a sensitivity analysis can then be estimated by
\[
\widehat{F}_k^{\overline{a}}(t;\Gamma) = n^{-1} \sum_{i=1}^{n} \int_{\xi} \frac{1}{(2\pi)^{1/2}} \exp\left(-\frac{\xi^2}{2}\right) \widehat\phi_k^{\overline{a}}(t;O_i,\xi,\Gamma) d\xi.
\]

\vspace{-0.4cm}
\section{Simulation}
\label{sec: simulation}
We evaluate the finite-sample performance of EIF-based estimation and inference through a simulation study for: (1) the type-1-specific counterfactual CIF; (2) the three decomposed effects (mediation effect, direct effect, and competing effect); and (3) the controlled counterfactual type-$k$-specific CIF, obtained by either (i) fixing the treatment and mediator level, or (ii) fixing the treatment, mediator level, and the hazard of the competing event. As described in detail in Supplementary Material F.1, data-generating processes satisfy Assumptions ~\ref{ass:std} and \ref{ass:nat}. We approximate the mediator density ratio with Riesz regression, estimate the hazard functions with semiparametric Cox regression, and estimate the propensity score with logistic regression.

Supplementary Material F.2 shows the simulation results. Supplementary Figures~S3 and~S4 show, for a single simulated dataset, the estimated (i) counterfactual CIFs, (ii) decomposed effects, and (iii) counterfactual risk as a function of the mediator level. Tables S1--S4 summarize the bias, mean EIF-based standard error, Monte Carlo standard deviation of the sampling distribution, and coverage probability of the proposed EIF-based estimators across multiple time points $t$ under various settings. Overall, the simulation results indicate that bias is generally small even at a sample size as small as $n=1000$. Moreover, the EIF-based standard error estimator closely approximates the Monte Carlo standard deviation of the sampling distribution, the empirical coverage of the 95\% confidence interval is close to the nominal level, and estimator performance generally improves as sample size increases.

Finally, in Supplementary Material F.3, we (1) demonstrate the multiple robustness property of the proposed EIF-based estimators by summarizing estimators' bias under various misspecifications of the nuisance parameters; and (2) examine alternative approaches to estimating the mediator density ratio, including (i) approximating the numerator and denominator with normal densities, and (ii) re-expressing the density ratio via Bayes' rule followed by estimation using generalized linear models. Our simulation results suggest that directly approximating the density ratio using Riesz regression often yields smaller finite-sample bias than other methods.

\vspace{-0.5cm}
\section{Two immune correlates studies} 
\label{application}
\vspace{-0.3cm}
\subsection{SARS-CoV-2}
\label{subsec: case study covid}
We apply the method to a mock dataset designed to mimic an ongoing Project NextGen vaccine trial \citep{hofmeyer2024project}, an active-controlled randomized trial comparing a novel oral COVID-19 vaccine ($A=1$) with an already approved injectable COVID-19 vaccine ($A=0$). The trial supports a range of immune correlates questions. One may ask, for instance, to what extent the cellular response level measured 30 days post-vaccination mediates the relative effect (oral versus injectable) on asymptomatic first SARS-CoV-2 infection ($k=1$), treating symptomatic first infection as a competing event. Our mock dataset mimics the summary statistics of the COV2008 trial, one of the most recent vaccine trials to whose data our group has access; see the Supplementary Material for details. The mock dataset includes $5000$ participants. The marker of interest is Day 30 BA.1/BA.2 spike peptide–stimulated CD4+ T cells expressing IFN-$\gamma$ and/or IL-2. About a quarter of participants in the dataset have their marker level measured under a two-phase sampling scheme.

\begin{figure}[!tbh]
\centering
\includegraphics[width=0.49\textwidth]{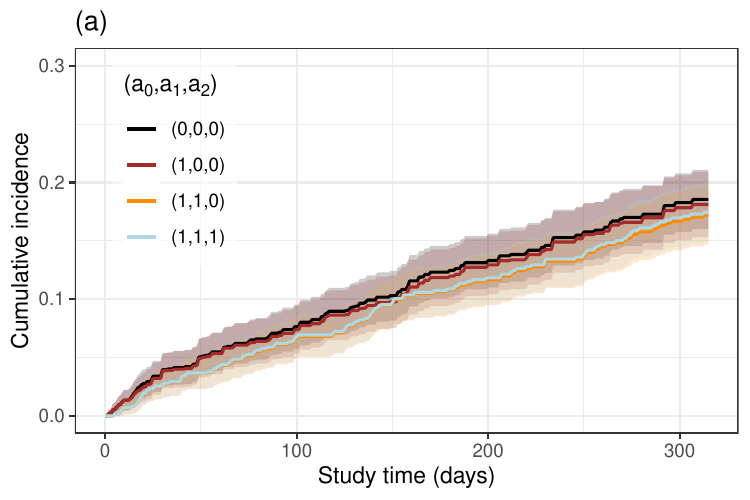} 
\includegraphics[width=0.49\textwidth]{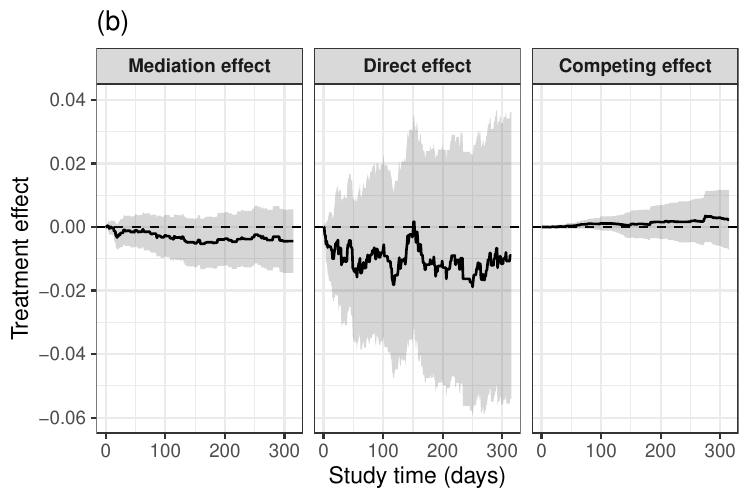} \\
\includegraphics[width=0.49\textwidth]{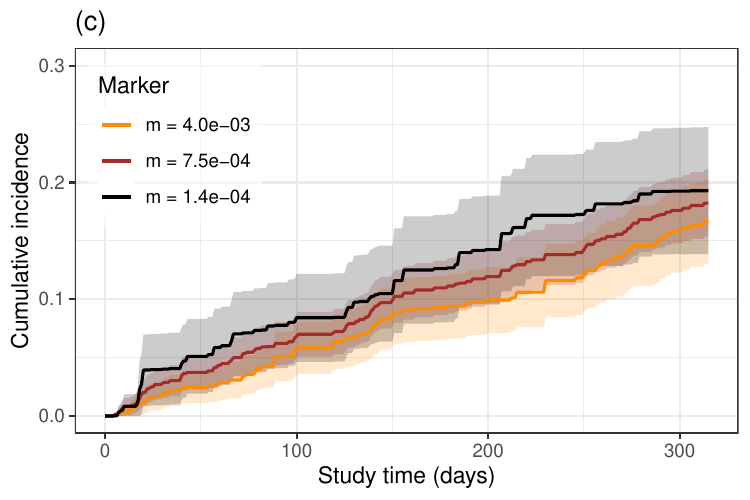} 
\includegraphics[width=0.49\textwidth]{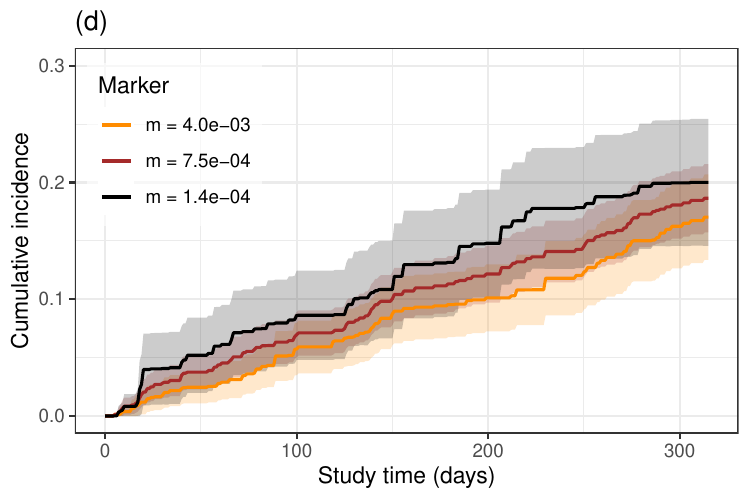} 
\caption{ {\small COVID-19 vaccine case study. (a) Estimated counterfactual CIFs of asymptomatic first infection under $\overline{a} = (0,0,0)$, $(1,0,0)$, $(1,1,0)$, and $(1,1,1)$. The first, second, and third components of $\overline{a}$ denote the treatment levels that generate the marker, asymptomatic first infection, and symptomatic first infection, respectively. (b) Estimated mediation effect, direct effect, and competing effect. (c)--(d) Estimated kernel-smoothed controlled CIFs of asymptomatic first infection under active treatment, with the Day 30 cellular response fixed at each of three levels. In (c), symptomatic first infection develops naturally under active treatment; in (d), its hazard under active treatment is halved.}} 
\label{covid2008_asymp}
\end{figure}

Figure~\ref{covid2008_asymp}(a) plots the estimated counterfactual cumulative incidence curves $\widehat{F}_1^{\overline{a}}(t)$ for the asymptomatic first infection, for $\overline{a} = (0,0,0)$, $(1,0,0)$, $(1,1,0)$, and $(1,1,1)$. Under the natural effects framework, $\widehat{F}_1^{(1,0,0)}(t)$ is the counterfactual cumulative incidence of asymptomatic first infection in a cross-world scenario in which the cellular response develops naturally under active treatment, while both asymptomatic and symptomatic infection processes develop naturally under control. As discussed previously, this same curve admits different interpretations under the stochastic interventional and separable effects frameworks. 

Figure~\ref{covid2008_asymp}(b) plots the three decomposed effects: (NME, NDE, NCE) under the natural effects framework, (IME, IDE, ICE) under the stochastic interventional effects framework (when the interventional distributions equal the natural ones), and (SME, SDE, SCE) under the separable effects framework. The estimated mediation effect is slightly negative but insignificant, consistent with the small treatment-induced shift in the Day 30 cellular response level in this mock dataset. The estimated direct effect is negative but small and not statistically significant, suggesting that the active treatment may lower the incidence of asymptomatic first infection through pathways that operate neither via the cellular response of interest nor via the competing event, such as humoral immunity. The estimated competing effect is small, positive, and not statistically significant, suggesting that the active treatment, compared with the control, may slightly reduce the incidence of symptomatic first infection, leaving more participants at risk of asymptomatic first infection and thereby slightly increasing its cumulative incidence. Based on the point estimate of the reduction in the CIF of asymptomatic first infection by Day 315 (0.011, SE 0.022), the direct effect accounts for 80.8\%, the mediation effect for 40.1\%, and the competing effect for $-20.9\%$. 

Figure~S11 in Section~G.1.3 of the Supplementary Material presents results under the alternative decomposition discussed after Theorem~\ref{thm:nat}. These results are qualitatively similar to those of the main analysis; the estimated interaction $I(t)$ is no larger than 0.0043. We further apply the proposed sensitivity analysis to assess how the direct effect and the competing indirect effect vary in the presence of an unmeasured frailty between the asymptomatic and symptomatic first infection processes; see Figure~S10 in Supplementary Material~G.1.2. 

%plots the two effects against the sensitivity parameter $\Gamma$. 

%The direct effect remains insignificant even when the hazard ratio of frailty is $\exp(1)$.

Figure~\ref{covid2008_asymp}(c) plots the estimated kernel-smoothed controlled counterfactual CIF for asymptomatic first infection, setting the treatment to the active treatment, fixing the Day 30 cellular response level at each of three levels, and letting the symptomatic first infection develop naturally under active treatment. These curves suggest that a higher cellular response level is associated with a higher incidence of asymptomatic first infection.
Figure~\ref{covid2008_asymp}(d) repeats this analysis with the hazard of a symptomatic first infection halved. As expected, reducing the symptomatic hazard raises the incidence of asymptomatic first infection, since more participants remain in the at-risk set. The increase is small, however, mainly because the rate of symptomatic first infection is low in the dataset to begin with.

Figure~S12 in the Supplementary Material~G.1.4 provides another way to visualize the mediator--outcome relationship by plotting the Day 315 cumulative incidence of asymptomatic first infection as a function of the cellular response level. 
Finally, Supplementary Material~G.1.5 repeats the analyses above with a symptomatic first infection as the event of interest and asymptomatic first infection as the competing event.

\vspace{-0.2cm}
\subsection{The Antibody Mediated Prevention trials}
\label{subsec: case study HIV}
The Antibody Mediated Prevention (AMP) trials (HVTN 703/HPTN 081 and HVTN 704/HPTN 085) were a pair of phase 2b, randomized, placebo-controlled studies testing whether VRC01, a broadly neutralizing monoclonal antibody (bnAb) targeting the CD4-binding site, could prevent HIV-1 acquisition when infused every eight weeks in HIV-uninfected individuals. Competing risks failure time analysis showed that VRC01 prevented HIV-1 acquisition events with VRC01-sensitive strains but did not prevent acquisition of resistant strains \citep{corey2021two}.

We apply the proposed controlled effects methods to study the marker--outcome relationship between the predicted Day 61 (D61) 80\% inhibitory dilution titer (predicted $\mathrm{ID}_{80}$) following the second infusion and acquisition of a VRC01-sensitive virus (defined as $\mathrm{IC}_{80} < 3\,\mu\mathrm{g}/\mathrm{mL}$), treating acquisition of a VRC01-insensitive virus as the competing event. For a given viral strain, the strain-specific predicted D61 $\mathrm{ID}_{80}$ is the VRC01 serum concentration measured at the visit 5 days after the second infusion, divided by that strain's $\mathrm{IC}_{80}$. These strain-specific values are then averaged over the panel of viruses isolated from all placebo cases, yielding the final ``predicted D61 $\mathrm{ID}_{80}$.'' The predicted D61 $\mathrm{ID}_{80}$ is zero for all placebo recipients, since they never received VRC01. The positivity assumption for the mediator density therefore fails, and the decomposed effects are not identified. Nevertheless, we can estimate the counterfactual risk of acquiring the virus under treatment given specific levels of predicted D61 $\mathrm{ID}_{80}$.

\begin{figure}[ht]
\centering
\includegraphics[width=0.49\textwidth]{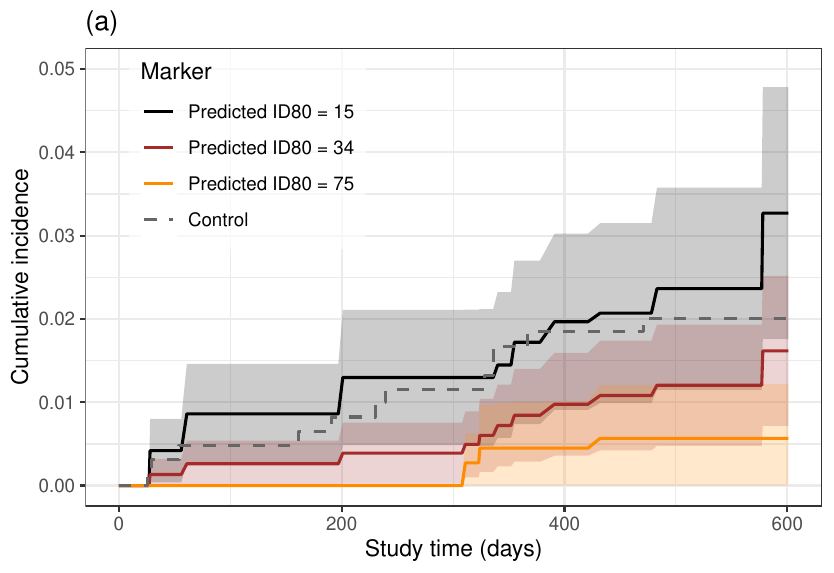} 
\includegraphics[width=0.49\textwidth]{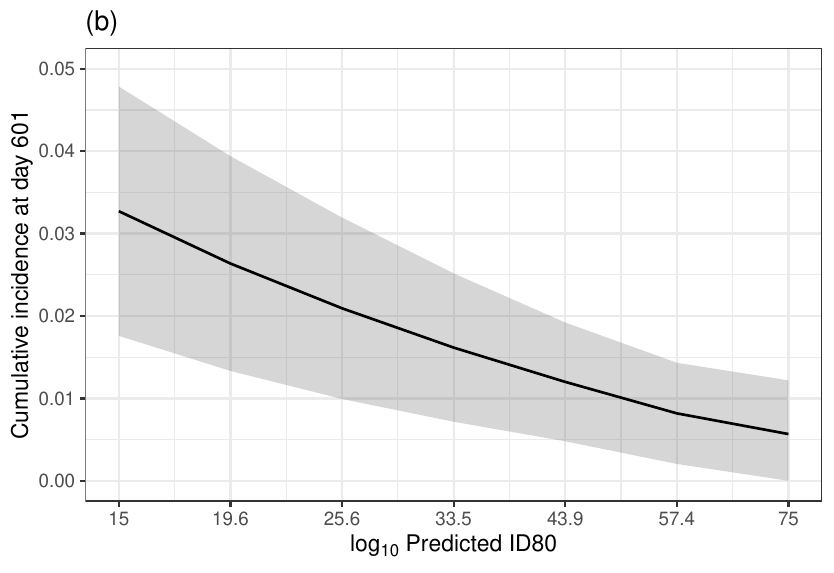} 
\caption{{\small AMP trial case study. (a) Estimated kernel-smoothed controlled counterfactual CIFs of HIV-1 acquisition with a VRC01-sensitive viral strain under assignment to VRC01, with the predicted D61 $\mathrm{ID}_{80}$ fixed at 15, 34, or 75. Dashed line: observed CIF of HIV-1 acquisition with a VRC01-sensitive viral strain in the placebo arm. (b) Estimated kernel-smoothed controlled counterfactual CIF at Day 601 of HIV-1 acquisition with a VRC01-sensitive strain, under assignment to VRC01, as a function of the predicted D61 $\mathrm{ID}_{80}$,} letting HIV-1 acquisition with a VRC01-insensitive viral strain occur naturally.} 
\label{amp_sensitive}
\end{figure}

Figure~\ref{amp_sensitive}(a) plots the estimated kernel-smoothed controlled counterfactual CIFs of HIV-1 acquisition with a VRC01-sensitive strain among VRC01 recipients, with the predicted D61 $\mathrm{ID}_{80}$ fixed at 15, 34, and 75, respectively, and acquisition with a VRC01-insensitive strain allowed to occur naturally. The dashed line represents the observed CIF in the placebo arm. The marker-specific efficacy by Day 601, defined as one minus the ratio of the marker-specific Day 601 risk to the placebo-arm Day 601 risk, against the VRC01-sensitive strain was estimated to be 71.8\% (95\% CI: $-1.1\%$ to $92.1\%$; $p = 0.052$) when the predicted D61 $\mathrm{ID}_{80}$ equals 75. Figure~\ref{amp_sensitive}(b) plots the estimated Day 601 cumulative incidence of HIV-1 acquisition with a VRC01-sensitive strain against the predicted D61 $\mathrm{ID}_{80}$, allowing acquisition with a VRC01-insensitive strain to occur naturally.

\vspace{-1cm}
\section{Discussion}
\label{sec: discussion}
We study causal mediation analysis for a time-to-event endpoint in the presence of competing events, addressing two tasks. First, we decompose the total effect of treatment on the type-specific cumulative incidence into an effect through the mediator, an effect through the competing events, and a direct effect, under three frameworks: natural effects, stochastic interventional effects, and separable effects. Second, we characterize how the type-specific cumulative incidence varies with the mediator level, either leaving the competing-event hazards at their natural levels or fixing them at levels of interest. Both tasks are central to a range of scientific questions, and for each we derive multiply robust estimators based on the efficient influence function.

Separating the indirect effect through the mediator from the total effect, under the natural and stochastic interventional effects frameworks, requires ruling out unmeasured mediator--outcome confounding. When a negative control outcome is available, such confounding may be detectable from its association with the mediator \citep{ashby2026validating}, under the mild assumption that a confounder of the mediator and the negative control is likely also a confounder of the mediator and the primary outcome. Under additional assumptions, it may further be corrected for; see, e.g., \citet{dukes2023proximal} and \citet{ashby2025debiasing}. Separating the indirect effect through competing events requires the additional assumption of no unmeasured common causes between the survival outcomes, which our sensitivity analysis examines.

In our effect decomposition, the direct and competing effects are formulated with respect to a common mediator density. To separate the effect of the mediator on each event, it may also be of interest to consider a hypothetical scenario in which event hazards depend on different mediator-generating processes. Under the stochastic interventional effects framework, we can assign a marginalized hazard for competing events $\{\lambda_j^*(t;a_j,x): j=1,\ldots,J,~j\neq k\}$ over the natural mediator distribution under treatment $a_j$ \citep{deng2026randomized}. However, this decomposition cannot be interpreted as a causal mechanism and is not sharp.
%Other strategies for effect decomposition are available. For example, one may decompose the total effect into a fraction that shifts the all-cause cumulative incidence due to treatment, a fraction that shifts the all-cause cumulative incidence due to the mediator, and a fraction that shifts the relative risk of the primary endpoint over competing events due to treatment. 

The natural and separable effects frameworks arrive at the same statistical functional (Theorems~\ref{thm:nat} and \ref{thm:sep}), so estimation, inference, and the robustness and efficiency properties developed in Section~\ref{sec: estimation and inference} apply to both without modification. This agreement should not be read as an equivalence of the frameworks: it is a consequence of the setting studied here --- a binary treatment and a single mediator measured once, shortly after treatment --- and does not hold in general. The two frameworks target genuinely different quantities under logically independent assumptions: cross-world independence can hold while treatment components are not dismissible, and vice versa. With longitudinal mediators, treatment-induced confounding of the mediator, or time-varying treatment, the natural effects functional is no longer identified, while the separable effects functional can be \citep{stensrud2022separable}. Even where the functionals coincide, the choice between frameworks remains a choice of interpretation and of which assumption one is prepared to defend, with the separable framework offering the advantage that its assumption is falsifiable in principle by a future trial that assigns treatment components independently.

The natural effects and stochastic interventional effects frameworks are inherently interventionist, in the sense that the mediator can be intervened upon, at least conceptually. This is appropriate for the immunological biomarker application we studied in this article. The separable effects framework is also interventionist, but operates under its own conceptualization of treatment components. Principal stratification offers a non-interventionist framework. Under the PS framework, and in the absence of competing events, one parameter of interest is the principal effect, which measures the effect of treatment on the latent population defined by the joint potential mediator levels \((M^1, M^0)\); see, e.g., \citet{follmann2006augmented} and \citet{gilbert2008evaluating} for applications to vaccine biomarker studies. In the presence of competing events, the definition of PS could be broadened, as the competing events offer an alternative basis for defining principal strata. For instance, in a vaccine study, \citet{rogawski2025vaccine} defined principal strata by a participant's first infection status and type of first infection (symptomatic or asymptomatic) under vaccination and control, and studied the vaccine's effect across these strata. An interesting future direction for mediation analysis would be to unify these different ways of defining principal strata --- those based on the mediator's joint potential values and those based on the joint potential values of the survival events --- into a single coherent framework, and to study which principal effects are identified under what assumptions.

\vspace{-0.5cm}
\section*{Data availability statement}
The AMP trials data are publicly available through Harvard Dataverse at
\url{https://doi.org/10.7910/DVN/WEJPUF}. 
%Access to these data may be available from the corresponding author or the original data custodians upon reasonable request and subject to the relevant institutional approvals and data-use agreements.

\vspace{-0.5cm}
\section*{Acknowledgements}
We thank the participants, site staff, and investigators of the Antibody Mediated Prevention trials sponsored by the National Institute of Allergy and Infectious Diseases of the National Institutes of Health (NIAID NIH). This work is funded by the NIAID NIH (R01AI192632, R37AI054165, R21AI190571, UM1AI068635 and UM1AI068614). The content is solely the responsibility of the authors and does not necessarily represent the official views of the NIH.

\vspace{-0.5cm}
\section*{Supplementary material}
The online supplementary material includes proofs of theoretical results (A--D), extensions (E), simulation (F), and additional data-analysis results (G). The \textsf{R} code implementing the proposed methods is available on GitHub: \url{https://github.com/naiiife/MedCmprsk}.

\singlespacing
\bibliographystyle{apalike}
\bibliography{ref}

@article{imai2010identification,
  title={Identification, inference and sensitivity analysis for causal mediation effects},
  author={Imai, Kosuke and Keele, Luke and Yamamoto, Teppei},
  journal={Statistical Science},
  volume={25},
  number={1},
  pages={51--71},
  year={2010}
}

@article{vanderweele2014effect,
  title={Effect decomposition in the presence of an exposure-induced mediator-outcome confounder},
  author={VanderWeele, Tyler J and Vansteelandt, Stijn and Robins, James M},
  journal={Epidemiology},
  volume={25},
  number={2},
  pages={300--306},
  year={2014},
  publisher={LWW}
}

@article{vansteelandt2017interventional,
  title={Interventional effects for mediation analysis with multiple mediators},
  author={Vansteelandt, Stijn and Daniel, Rhian M},
  journal={Epidemiology},
  volume={28},
  number={2},
  pages={258--265},
  year={2017},
  publisher={LWW}
}

@article{diaz2020causal,
  title={Causal mediation analysis for stochastic interventions},
  author={D{\'\i}az, Iv{\'a}n and Hejazi, Nima S},
  journal={Journal of the Royal Statistical Society Series B: Statistical Methodology},
  volume={82},
  number={3},
  pages={661--683},
  year={2020},
  publisher={Oxford University Press}
}

@article{miles2020semiparametric,
  title={On semiparametric estimation of a path-specific effect in the presence of mediator-outcome confounding},
  author={Miles, Caleb H and Shpitser, Ilya and Kanki, Phyllis and Meloni, Seema and {Tchetgen Tchetgen}, Eric J},
  journal={Biometrika},
  volume={107},
  number={1},
  pages={159--172},
  year={2020},
  publisher={Oxford University Press}
}

@incollection{robins2022interventionist,
  title={An interventionist approach to mediation analysis},
  author={Robins, James M and Richardson, Thomas S and Shpitser, Ilya},
  booktitle={Probabilistic and causal inference: the works of Judea Pearl},
  pages={713--764},
  year={2022},
  publisher={Oxford University Press}
}

@article{Vo2020Longitudinal,
title={Longitudinal mediation analysis of time-to-event endpoints in the presence of competing risks},
author={Tat-Thang Vo and Hilary Davies-Kershaw and R. Hackett and S. Vansteelandt},
journal={Lifetime Data Analysis},
year={2020},
volume={28},
pages={380 - 400}
}

@article{frangakis2002principal,
  title={Principal stratification in causal inference},
  author={Frangakis, Constantine E and Rubin, Donald B},
  journal={Biometrics},
  volume={58},
  number={1},
  pages={21--29},
  year={2002},
  publisher={Oxford University Press}
}

@article{ashby2025debiasing,
  title={Debiasing hazard-based, time-varying vaccine effects using vaccine-irrelevant infections: {An} observational extension of a pivotal {Phase 3 COVID-19} vaccine efficacy trial},
  author={Ashby, Ethan and Follmann, Dean and Janes, Holly and Gilbert, Peter B and Ye, Ting and Baden, Lindsey R and Sahly, Hana M El and Zhang, Bo},
  journal={arXiv preprint arXiv:2511.15099},
  year={2025}
}

@article{dukes2023proximal,
  title={Proximal mediation analysis},
  author={Dukes, Oliver and Shpitser, Ilya and Tchetgen Tchetgen, Eric J},
  journal={Biometrika},
  volume={110},
  number={4},
  pages={973--987},
  year={2023},
  publisher={Oxford University Press}
}

@article{Diaz2022Causal,
title={Causal survival analysis under competing risks using longitudinal modified treatment policies},
author={Iv{\'a}n D{\'\i}az and K. Hoffman and N. Hejazi},
journal={Lifetime Data Analysis},
year={2022},
volume={30},
pages={213-236}
}

@article{lange2011direct,
  title={Direct and indirect effects in a survival context},
  author={Lange, Theis and Hansen, J{\o}rgen V},
  journal={Epidemiology},
  volume={22},
  number={4},
  pages={575--581},
  year={2011}
}

@article{tchetgen2011causal,
  title={On causal mediation analysis with a survival outcome},
  author={Tchetgen Tchetgen, Eric J},
  journal={The International Journal of Biostatistics},
  volume={7},
  number={1},
  year={2011},
  publisher={De Gruyter}
}

@article{didelez2019defining,
  title={Defining causal mediation with a longitudinal mediator and a survival outcome},
  author={Didelez, Vanessa},
  journal={Lifetime Data Analysis},
  volume={25},
  pages={593--610},
  year={2019},
  publisher={Springer}
}

@article{huang2021causal,
  title={Causal mediation of semicompeting risks},
  author={Huang, Yen-Tsung},
  journal={Biometrics},
  volume={77},
  number={4},
  pages={1143--1154},
  year={2021},
  publisher={Wiley Online Library}
}

@article{deng2024direct,
  title={Direct and indirect treatment effects in the presence of semicompeting risks},
  author={Deng, Yuhao and Wang, Yi and Zhou, Xiao-Hua},
  journal={Biometrics},
  volume={80},
  number={2},
  pages={ujae032},
  year={2024},
  publisher={Oxford University Press}
}

@article{stensrud2021generalized,
  title={A generalized theory of separable effects in competing event settings},
  author={Stensrud, Mats J and Hern{\'a}n, Miguel A and {Tchetgen Tchetgen}, Eric J and Robins, James M and Didelez, Vanessa and Young, Jessica G},
  journal={Lifetime Data Analysis},
  volume={27},
  number={4},
  pages={588--631},
  year={2021},
  publisher={Springer}
}

@article{stensrud2022separable,
  title={Separable effects for causal inference in the presence of competing events},
  author={Stensrud, Mats J and Young, Jessica G and Didelez, Vanessa and Robins, James M and Hern{\'a}n, Miguel A},
  journal={Journal of the American Statistical Association},
  volume={117},
  number={537},
  pages={175--183},
  year={2022},
  publisher={Taylor \& Francis}
}

@article{gilbert2023controlled,
    title={A controlled effects approach to assessing immune correlates of protection},
    author={Gilbert, Peter B and Fong, Youyi and Kenny, Avi and Carone, Marco},
    journal={Biostatistics},
    volume={24},
    number={4},
    pages={850--865},
    year={2023}
}

@article{joffe2009related,
  title={Related causal frameworks for surrogate outcomes},
  author={Joffe, Marshall M and Greene, Tom},
  journal={Biometrics},
  volume={65},
  number={2},
  pages={530--538},
  year={2009},
  publisher={Oxford University Press}
}

@inproceedings{richardson2013single,
  title={Single world intervention graphs: a primer},
  author={Richardson, Thomas S and Robins, James M},
  booktitle={Second UAI workshop on causal structure learning, Bellevue, Washington},
  year={2013}
}

@article{breum2024estimation,
  title={Estimation of separable direct and indirect effects in a continuous-time illness-death model},
  author={Breum, Marie Skov and Munch, Anders and Gerds, Thomas A and Martinussen, Torben},
  journal={Lifetime Data Analysis},
  volume={30},
  number={1},
  pages={143--180},
  year={2024},
  publisher={Springer}
}

@article{martinussen2021estimation,
  title={Estimation of separable direct and indirect effects in continuous time},
  author={Martinussen, Torben and Stensrud, Mats Julius},
  journal={Biometrics},
  volume={79},
  number={1},
  pages={127-139},
  year={2023},
  publisher={Wiley Online Library}
}

@incollection{pearl2022direct,
  title={Direct and indirect effects},
  author={Pearl, Judea},
  booktitle={Probabilistic and causal inference: the works of Judea Pearl},
  pages={373--392},
  year={2022}
}

@article{vanderweele2008simple,
  title={Simple relations between principal stratification and direct and indirect effects},
  author={VanderWeele, Tyler J},
  journal={Statistics \& Probability Letters},
  volume={78},
  number={17},
  pages={2957--2962},
  year={2008},
  publisher={Elsevier}
}

@article{follmann2006augmented,
  title={Augmented designs to assess immune response in vaccine trials},
  author={Follmann, Dean},
  journal={Biometrics},
  volume={62},
  number={4},
  pages={1161--1169},
  year={2006},
  publisher={Oxford University Press}
}

@article{rose2011targeted,
  title={A targeted maximum likelihood estimator for two-stage designs},
  author={Rose, Sherri and van der Laan, Mark J},
  journal={The International Journal of Biostatistics},
  volume={7},
  number={1},
  pages={17},
  year={2011}
}

@article{domingo2026path,
  title={A Path-Specific Effect Approach to Mediation Analysis With Time-Varying Mediators and Time-to-Event Outcomes Accounting for Competing Risks},
  author={Domingo-Relloso, Arce and Zhang, Yuchen and Wang, Ziqing and Suchy-Dicey, Astrid M and Buchwald, Dedra S and Navas-Acien, Ana and Schwartz, Joel and Berhane, Kiros and Coull, Brent A and Valeri, Linda},
  journal={Statistics in Medicine},
  volume={45},
  number={3-5},
  pages={e70425},
  year={2026},
  publisher={Wiley Online Library}
}

@article{deng2026randomized,
  title={Randomized Interventional Effects in Semicompeting Risks, With Application to a Hematopoietic Cell Transplantation Study},
  author={Deng, Yuhao and Wang, Rui and Zhang, Tao and Zhan, Xiang},
  journal={Statistics in Medicine},
  volume={45},
  number={13-14},
  pages={e70628},
  year={2026},
  publisher={Wiley Online Library}
}

@article{gilbert2022immune,
  title={Immune correlates analysis of the {mRNA-1273 COVID-19} vaccine efficacy clinical trial},
  author={Gilbert, Peter B and Montefiori, David C and McDermott, Adrian B and Fong, Youyi and Benkeser, David and Deng, Weiping and Zhou, Honghong and Houchens, Christopher R and Martins, Karen and Jayashankar, Lakshmi and others},
  journal={Science},
  volume={375},
  number={6576},
  pages={43--50},
  year={2022},
  publisher={American Association for the Advancement of Science}
}

@article{kennedy2017non,
  title={Non-parametric methods for doubly robust estimation of continuous treatment effects},
  author={Kennedy, Edward H and Ma, Zongming and McHugh, Matthew D and Small, Dylan S},
  journal={Journal of the Royal Statistical Society Series B: Statistical Methodology},
  volume={79},
  number={4},
  pages={1229--1245},
  year={2017},
  publisher={Oxford University Press}
}

@article{gallop2009mediation,
  title={Mediation analysis with principal stratification},
  author={Gallop, Robert and Small, Dylan S and Lin, Julia Y and Elliott, Michael R and Joffe, Marshall and Ten Have, Thomas R},
  journal={Statistics in Medicine},
  volume={28},
  number={7},
  pages={1108--1130},
  year={2009},
  publisher={Wiley Online Library}
}

@article{hofmeyer2024project,
  title={{Project NextGen}: developing the next generation of {COVID-19} vaccines and therapeutics to respond to the present and prepare for the future},
  author={Hofmeyer, Kimberly A and Ventura, Christy L and Armstrong, Kimberly L and Houchens, Christopher R and Patel, Sandeep and Disbrow, Gary L and Johnson, Robert A and Bandremer, Aaron C and Beigel, John H and Belski, Timothy T and Ruben DO and Jayashankar, Lakshmi and Koup, Richard A and Kovacs, Gerald R and Link, Malen A and Martins, Karen A and Mason, Robin M and Stronsky, Sabrina M and Wolfe, Daniel N},
  journal={Clinical Infectious Diseases},
  volume={79},
  number={1},
  pages={115--121},
  year={2024},
  publisher={Oxford University Press US}
}

@article{ashby2026validating,
  title={Validating and leveraging {non-SARS-CoV-2} respiratory infection as a negative control outcome in a phase 3 {COVID-19} vaccine trial with extended observational follow-up},
  author={Ashby, Ethan and Janes, Holly and Follmann, Dean and Gilbert, Peter B and Zhou, Honghong and Wang, Xiaowei and Girard, Bethany and Priddy, Frances and Kublin, James G and Corey, Lawrence and others},
  journal={American Journal of Epidemiology},
  volume={195},
  number={1},
  pages={168--177},
  year={2026},
  publisher={Oxford University Press}
}

@article{corey2021two,
  title={Two randomized trials of neutralizing antibodies to prevent {HIV-1} acquisition},
  author={Corey, Lawrence and Gilbert, Peter B and Juraska, Michal and Montefiori, David C and Morris, Lynn and Karuna, Shelly T and Edupuganti, Srilatha and Mgodi, Nyaradzo M and Decamp, Allan C and Rudnicki, Erika and others},
  journal={New England Journal of Medicine},
  volume={384},
  number={11},
  pages={1003--1014},
  year={2021},
  publisher={Mass Medical Soc}
}

@article{rogawski2025vaccine,
  title={Vaccine efficacy against naturally asymptomatic infections: A novel estimand for quantifying vaccine effects},
  author={Rogawski McQuade, Elizabeth and Nabi, Razieh and Codi, Allison and Dean, Natalie and Lipsitch, Marc and Benkeser, David},
  journal={medRxiv},
  pages={2025--08},
  year={2025},
  publisher={Cold Spring Harbor Laboratory Press}
}

@article{robins1992identifiability,
  title={Identifiability and exchangeability for direct and indirect effects},
  author={Robins, James M and Greenland, Sander},
  journal={Epidemiology},
  volume={3},
  number={2},
  pages={143--155},
  year={1992},
  publisher={LWW}
}

@article{pearl2014interpretation,
  title={Interpretation and identification of causal mediation.},
  author={Pearl, Judea},
  journal={Psychological Methods},
  volume={19},
  number={4},
  pages={459},
  year={2014},
  publisher={American Psychological Association}
}

@article{gilbert2008evaluating,
  title={Evaluating candidate principal surrogate endpoints},
  author={Gilbert, Peter B and Hudgens, Michael G},
  journal={Biometrics},
  volume={64},
  number={4},
  pages={1146--1154},
  year={2008},
  publisher={Oxford University Press}
}

@article{branson2023causal,
  title={Causal effect estimation after propensity score trimming with continuous treatments},
  author={Branson, Zach and Kennedy, Edward H and Balakrishnan, Sivaraman and Wasserman, Larry},
  journal={arXiv preprint arXiv:2309.00706},
  year={2023}
}

@article{chernozhukov2021automatic,
  title={Automatic debiased machine learning via {Riesz} regression},
  author={Chernozhukov, Victor and Newey, Whitney K and Quintas-Martinez, Victor and Syrgkanis, Vasilis},
  journal={arXiv preprint arXiv:2104.14737},
  year={2021}
}

@article{chernozhukov2018double,
  title={Double/debiased machine learning for treatment and structural parameters},
  author={Chernozhukov, Victor and Chetverikov, Denis and Demirer, Mert and Duflo, Esther and Hansen, Christian and Newey, Whitney and Robins, James},
  journal={The Econometrics Journal},
  volume={21},
  issue={1},
  pages={C1-C68},
  year={2018}
}

@article{zhang2025neutralizing,
  title={{Neutralizing antibody immune correlates in COVAIL trial recipients of an mRNA second COVID-19 vaccine boost}},
  author={Zhang, Bo and Fong, Youyi and Dang, Lauren and Fintzi, Jonathan and Chen, Shiyu and Wang, Jing and Rouphael, Nadine G and Branche, Angela R and Diemert, David J and Falsey, Ann R and others},
  journal={Nature Communications},
  volume={16},
  number={1},
  pages={759},
  year={2025}
}

@article{zhang2024omicron,
  title={Omicron {COVID-19} immune correlates analysis of a third dose of {mRNA-1273} in the {COVE} trial},
  author={Zhang, Bo and Fong, Youyi and Fintzi, Jonathan and Chu, Eric and Janes, Holly E and Kenny, Avi and Carone, Marco and Benkeser, David and van der Laan, Lars WP and Deng, Weiping and others},
  journal={Nature Communications},
  volume={15},
  number={1},
  pages={7954},
  year={2024},
  publisher={Nature Publishing Group UK London}
}

@article{benkeser2021inference,
  title={Inference for natural mediation effects under case-cohort sampling with applications in identifying COVID-19 vaccine correlates of protection},
  author={Benkeser, David and D{\'\i}az, Iv{\'a}n and Ran, Jialu},
  journal={arXiv preprint arXiv:2103.02643},
  year={2021}
}

@article{baden2021efficacy,
  title={Efficacy and safety of the mRNA-1273 SARS-CoV-2 vaccine},
  author={Baden, Lindsey R and El Sahly, Hana M and Essink, Brandon and Kotloff, Karen and Frey, Sharon and Novak, Rick and Diemert, David and Spector, Stephen A and Rouphael, Nadine and Creech, C Buddy and others},
  journal={New England Journal of Medicine},
  volume={384},
  number={5},
  pages={403--416},
  year={2021},
  publisher={Mass Medical Soc}
}

@article{kanamori2009least,
  title={A least-squares approach to direct importance estimation},
  author={Kanamori, Takafumi and Hido, Shohei and Sugiyama, Masashi},
  journal={The Journal of Machine Learning Research},
  volume={10},
  pages={1391--1445},
  year={2009},
  publisher={JMLR. org}
}

@article{sugiyama2012density,
  title={Density-ratio matching under the {Bregman} divergence: a unified framework of density-ratio estimation},
  author={Sugiyama, Masashi and Suzuki, Taiji and Kanamori, Takafumi},
  journal={Annals of the Institute of Statistical Mathematics},
  volume={64},
  number={5},
  pages={1009--1044},
  year={2012},
  publisher={Springer}
}

@article{wager2018estimation,
  title={Estimation and inference of heterogeneous treatment effects using random forests},
  author={Wager, Stefan and Athey, Susan},
  journal={Journal of the American Statistical Association},
  volume={113},
  number={523},
  pages={1228--1242},
  year={2018},
  publisher={Taylor \& Francis}
}

\includepdf[pages=1-45]{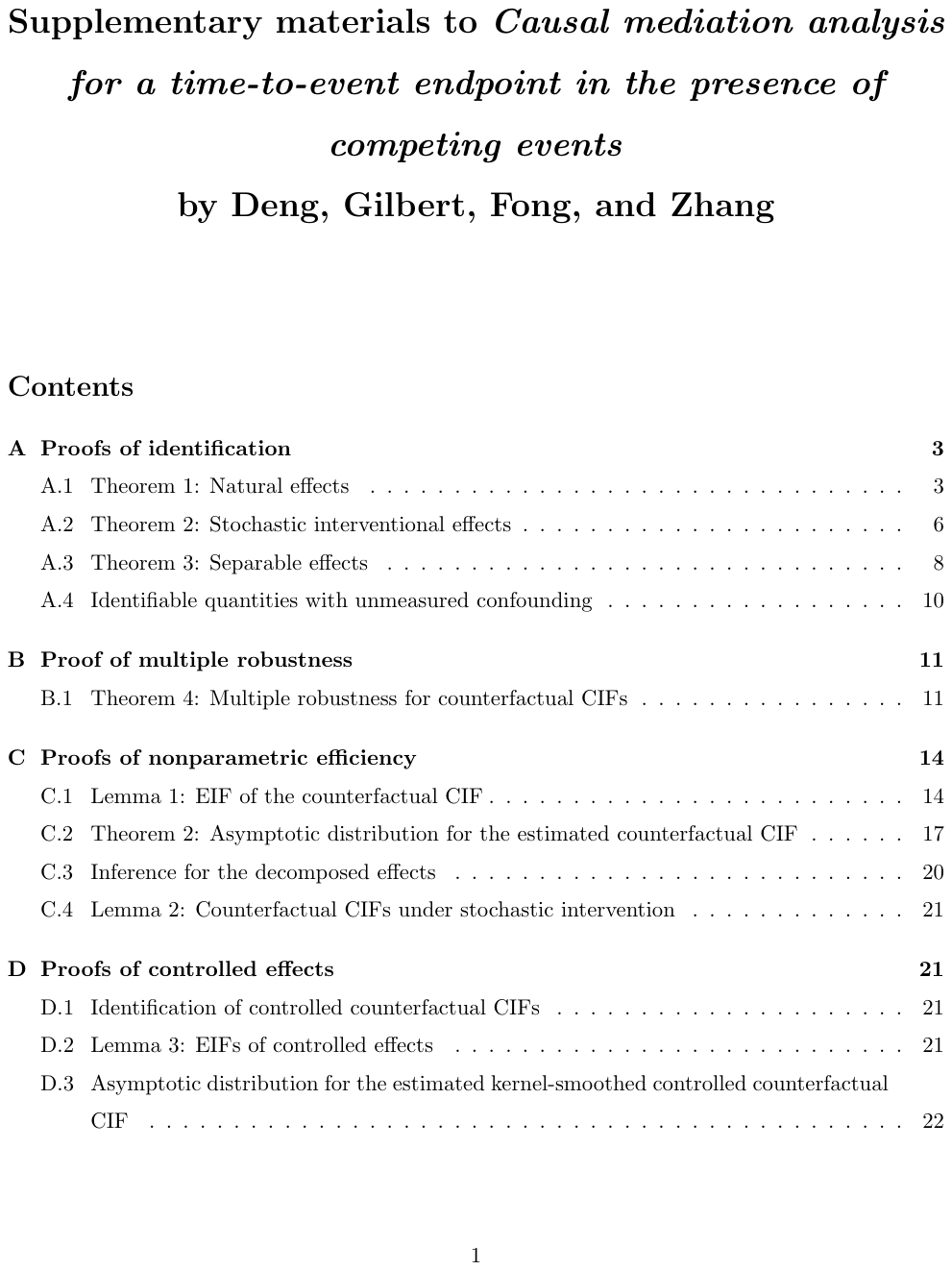}

\end{document}